\documentclass[a4paper,11pt,english]{article}
\pdfoutput=1
\usepackage{jinstpub} 
\usepackage[T1]{fontenc}
\usepackage{babel}
\usepackage{txfonts}
\usepackage{mathptmx}
\usepackage{lmodern}
\usepackage{blindtext}
\usepackage{amsmath}
\usepackage[utf8]{inputenc}
\usepackage{newunicodechar}
\usepackage{graphicx}% Include figure files 
\usepackage{comment}

\usepackage{subfigure}

\title{Single-hit resolution measurement \\ with MEG~II drift chamber prototypes}

\author[a]{A.M.~Baldini,}
\author[b]{E.~Baracchini,}
\author[c]{G.~Cavoto,}
\author[d,e]{M.~Cascella,}
\author[a,f]{F.~Cei,}
\author[a,f]{M.~Chiappini,}
\author[d,e]{G.~Chiarello,}
\author[d,e]{C.~Chiri,}
\author[a]{S.~Dussoni,}
\author[a]{L.~Galli,}
\author[e]{F.~Grancagnolo,}
\author[a]{M.~Grassi,}
\author[c,g]{V.~Martinelli,}
\author[a,f]{D.~Nicol\`o,}
\author[d,e]{M.~Panareo,}
\author[d,e]{A.~Pepino,}
\author[c]{G.~Piredda,}
\author[c]{F.~Renga,}
\author[c,g]{E.~Ripiccini,}
\author[a]{G.~Signorelli,}
\author[d,e]{G.F.~Tassielli,}
\author[a]{F.~Tenchini,}
\author[a,h,1]{M.~Venturini\note{Corresponding author.},}
\author[c]{C.~Voena.}

\affiliation[a]{INFN Sezione di Pisa, Largo B. Pontecorvo 3, 56127 Pisa, Italy.}
\affiliation[b]{ICEPP, University of Tokyo 7-3-1 Hongo, Bunkyo-ku, Tokyo 113-0033, Japan.}
\affiliation[c]{INFN Sezione di Roma, Piazzale A. Moro, 00185 Roma, Italy}
\affiliation[d]{Dipartimento di Matematica e Fisica, Universit\`a del Salento, Via per Arnesano, 73100 Lecce, Italy}
\affiliation[e]{INFN Sezione di Lecce, Via per Arnesano, 73100 Lecce, Italy}
\affiliation[f]{Dipartimento di Fisica, Universit\`a di Pisa, Largo B. Pontecorvo 3, 56127 Pisa, Italy.}
\affiliation[g]{Dipartimento di Fisica, Universit\`a ``Sapienza'', Piazzale A. Moro, 00185 Roma, Italy}
\affiliation[h]{Scuola Normale Superiore, Piazza dei Cavalieri 7, 56126 Pisa, Italy.}

\emailAdd{marco.venturini@pi.infn.it}

\abstract{
Drift chambers operated with helium-based gas mixtures represent a common solution for tracking charged particles keeping the material budget in the sensitive volume to a minimum.
The drawback of this solution is the worsening of the spatial resolution due to primary ionisation fluctuations, which is a limiting factor for high granularity drift chambers like the MEG~II tracker.
We report on the measurements performed on three different prototypes of the MEG~II drift chamber aimed at determining the achievable single-hit resolution.
The prototypes were operated with helium/isobutane gas mixtures and exposed to  cosmic rays, electron beams and radioactive sources. Direct measurements of the single hit resolution performed with an external tracker returned a value of 110~$\mu$m, consistent with the values obtained with indirect measurements performed with the other prototypes.
}

\keywords{ Gaseous detectors;
Wire chambers;
Particle tracking detectors;
Performance of High Energy Physics Detectors.
}

\begin{document}
\maketitle
\flushbottom

\section{Introduction}

The MEG~II experiment~\cite{bib:proposal} will search for the lepton-flavour-violating decay $\mu^+ \rightarrow e^+ \gamma$ with a Branching Ratio sensitivity down to $5\times 10^{-14}$. In this experiment, the trajectory of positrons emitted in muon decays will be reconstructed by a high transparency stereo drift chamber, operated with a gas mixture of helium and isobutane in order to have good momentum and angular resolution for low-momentum ($\approx$~50~MeV/$c$) particles.
The single-hit resolution of this chamber, together with the high granularity and the low multiple scattering contribution, plays a fundamental role in the overall sensitivity of the experiment, since it directly influences the positrons reconstructed kinematic variables.

The detector is a 2~m long, full stereo cylindrical drift chamber made of ten layers of approximately square cells at stereo angle of about $7^\circ$ with alternating signs. The cell width increases linearly with the radius of the layer and, because of the stereo angle, varies slightly with the position along the chamber axis ($z$~axis), ranging from 6.7~mm (inner layer, $z = 0$) to 9~mm (outer layer, maximum $z$). 
 The requirement of high transparency is fulfilled by using a 85\%--15\% helium-isobutane low mass gas mixture which however produces a low density of primary ionization clusters and limits the chamber spatial resolution at small impact parameters. This effect, particularly relevant for small cells, is however mitigated by the average polar angle (about $50^\circ$) of the positron trajectories, which results in a 30$\%$ increase of the total number of ionisation clusters with respect to normal incidence.

We report here about the single cell spatial resolution obtained with three different prototypes. 
The paper is organized as follows. The experimental set-ups for the three different prototypes are 
shown in section~\ref{sec:hardware}; section~\ref{sec:indirect} contains the description of the 
indirect measurements of single-hit resolution extracted from these, whereas in 
section~\ref{sec:direct} a direct measurement achieved with a high resolution external tracker is 
reported for one of the prototypes. Final conclusions are given in section~\ref{sec:conclusion}.

\section{The drift chamber prototypes}
\label{sec:hardware}
%%%%%%%%%%%%%%%%%%%%%%%%%%%%%%%%%%%%%%%%%%%%%%%%%%%%%%%%%%
%%%%%%%%%%%%%%%%%%%%%%%%%%% LE %%%%%%%%%%%%%%%%%%%%%%%%%%%
\subsection{Three-tubes prototype}
\label{sec:tritubo}
\begin{figure}[!b]
\centering
\subfigure[]{
\includegraphics[width=0.13\textwidth]{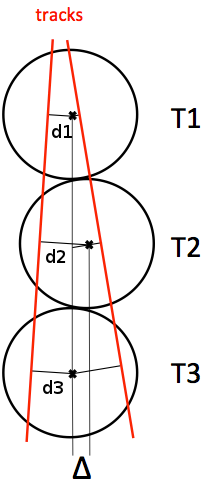}\label{fig:Tritube}}
\hspace{2cm}
\subfigure[]{\includegraphics[width=0.4\textwidth]{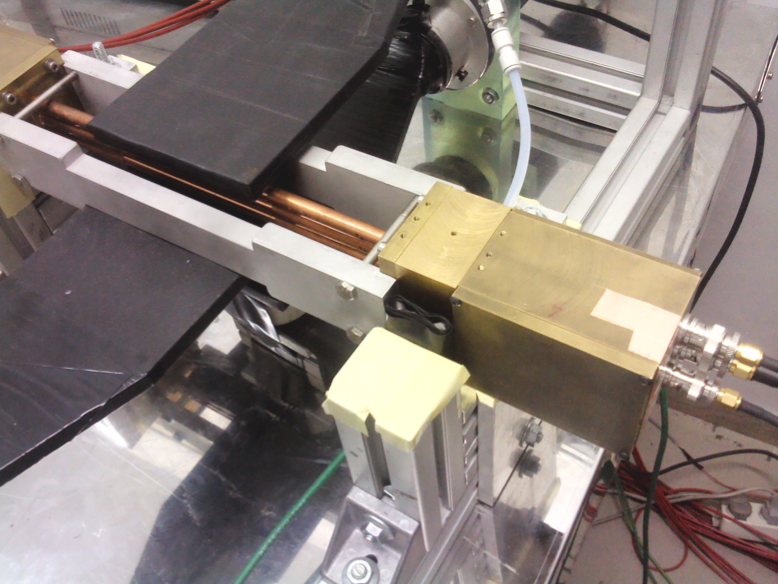}
\label{fig:Tritubepic}}
\protect\caption{(a) Definition of $d_{1},d_{2},d_{3}$, $\Delta$ with tracks examples. (b) The three-tubes set-up. }
\end{figure}
\begin{figure}[!t]
\begin{centering}
\includegraphics[width=0.8\textwidth]{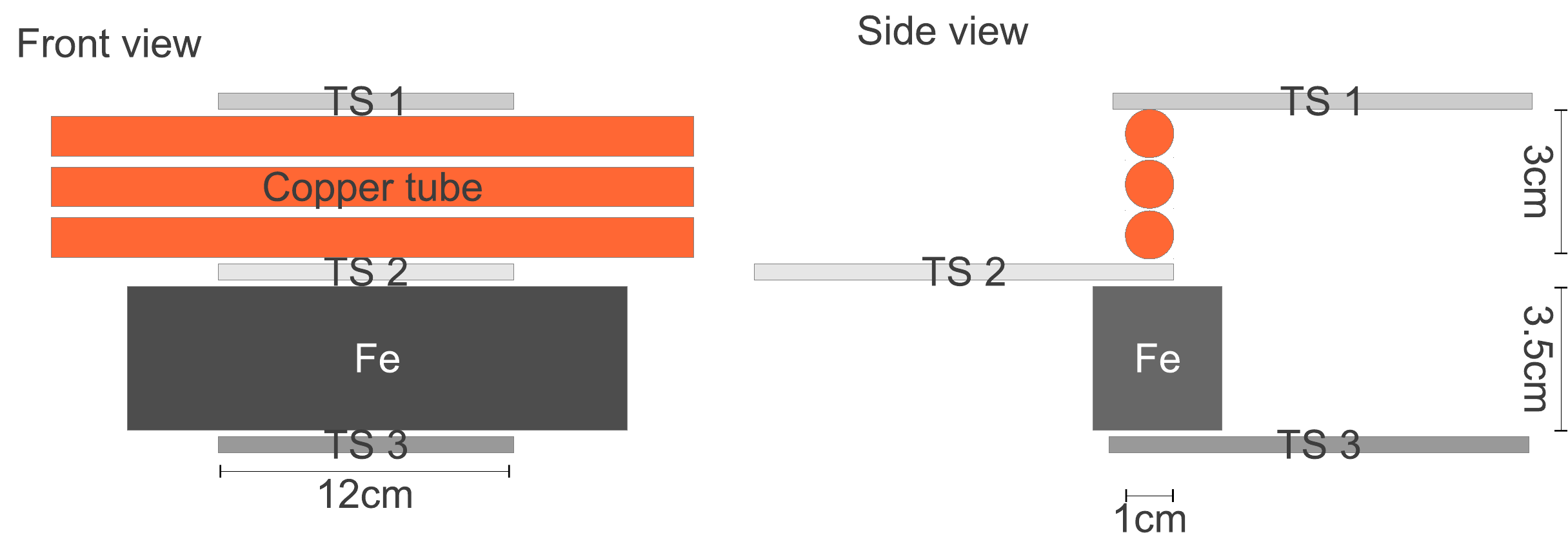}
\par\end{centering}
\protect\caption{\label{fig:tritube_scheme_2} Schematic view of the three-tubes system.}
\end{figure}

A system of three parallel copper drift tubes, having 8~mm internal diameter, 30~cm length and 500~$\mbox{\ensuremath{\mu}m}$ wall thickness, has been used for a first check of the performance of a drift cell similar to the one of the MEG II drift chamber. 
Anodes are 20~$\mbox{\ensuremath{\mu}m}$ gold-plated tungsten wires and are set at a high voltage of 1500~V.
The middle tube is staggered with respect to the outer ones by $\Delta=500\,\mbox{\ensuremath{\mu}m}$. A scheme of the apparatus is reported in figure~\ref{fig:Tritube}. 
The sense wires connections occur inside brass boxes  with gas tight connectors. The signals are read out by two Phillips 775 amplifiers, with a total gain of a factor~100 and a bandwidth of 1.8~GHz, and a Tektronix TDS7404 oscilloscope, with 4~GHz bandwidth. The set-up, shown in figure~\ref{fig:Tritubepic}, is completed by a cosmic ray trigger made by the coincidence of three plastic scintillators. Two scintillators are placed just above and below the tubes. The third scintillator is placed under a $3.5$~cm thick iron slab (see figure~\ref{fig:tritube_scheme_2}) in order to select tracks as vertical as possible and to remove the low-energy component of cosmic rays.

%%%%%%%%%%%%%%%%%%%%%%%%%%%%%%%%%%%%%%%%%%%%%%%%%%%%%%%%%%

\subsection{Three-cells prototype}
\label{sec:tricella}
\begin{figure}[!b]
  \centering
 \subfigure[]{ \includegraphics[height=.3\textwidth, angle=0]{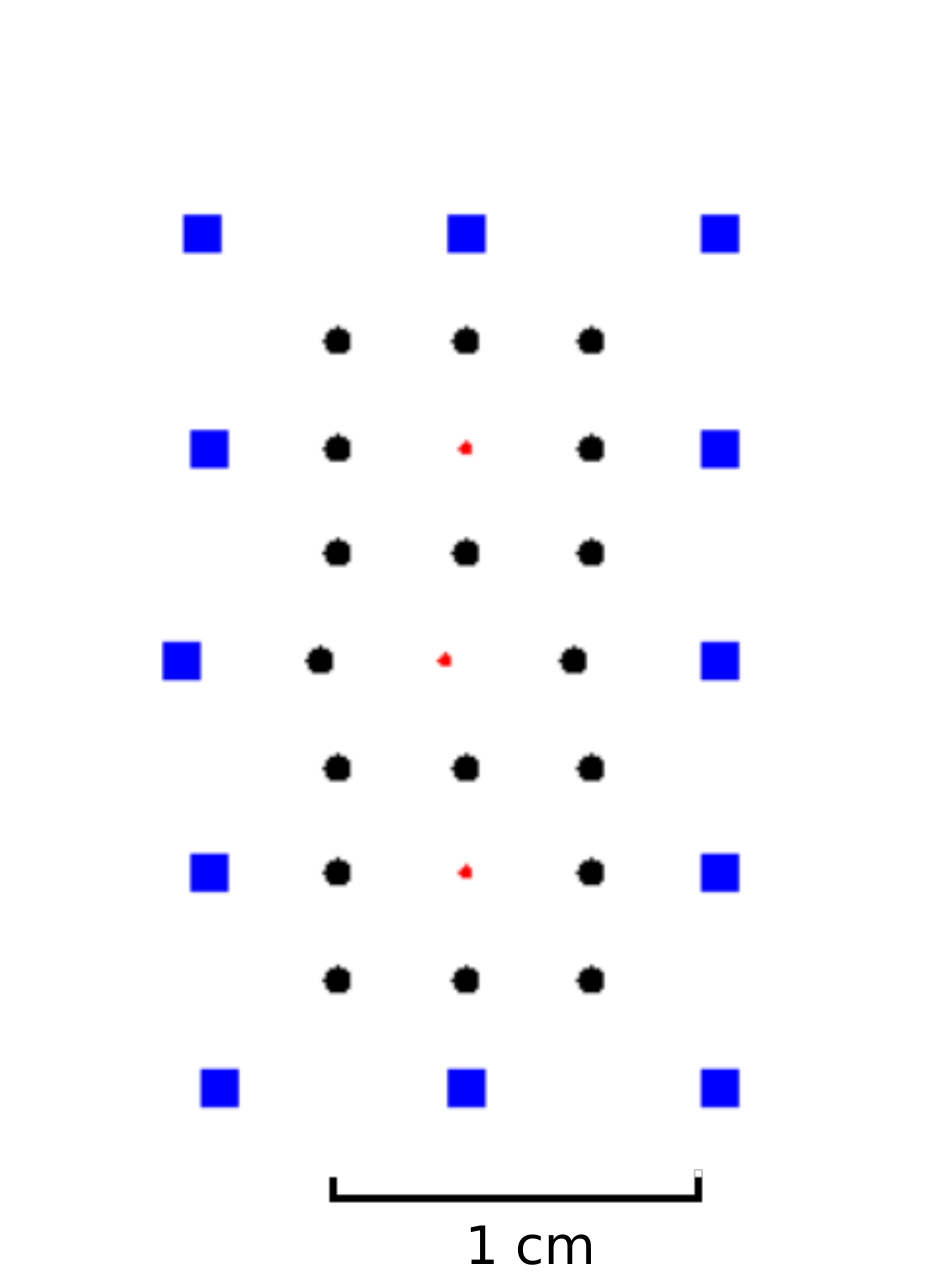}\label{TriCell} }
\subfigure[]{ \includegraphics[ trim=0 10cm 0 10cm, clip=true,height=.25\textwidth]{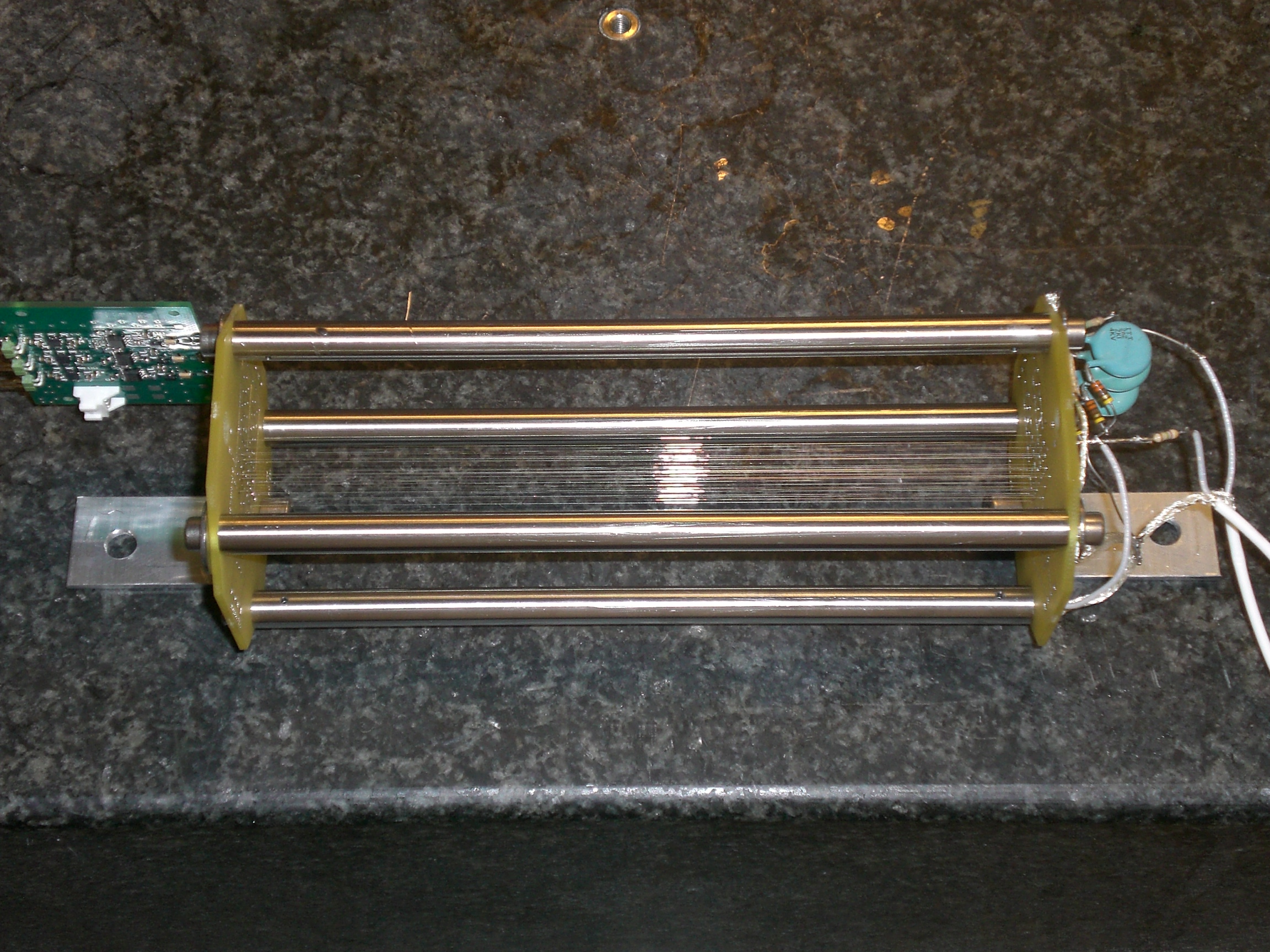} \label{tritubonobox}}  
\caption{(a) Scale schematics of the three-cells configuration with anode wires (small red circles), cathode wires (big black circles) and guard wires (blue squares).
(b) Picture of three-cell prototype: circuitry for power supply is visible on the right side,
while the three-channel pre-amplifier is connected on the left side. }

\end{figure}

The second prototype has a square cell geometry and it is composed of three adjacent drift cells.
The wire pattern of three 7 mm squared cells is reproduced on two Printed Circuit Boards (PCBs), as shown in figure~\ref{TriCell}, in order to simulate, at a given $z$, shape and dimensions of the MEG II drift chamber cell. As in the three-tubes layout, the anode wire of the central cell is staggered by 500~$\mu$m. 
  The wires used are 20-$\mu$m gold-plated tungsten anodes and 80-$\mu$m silver-plated aluminium cathodes and guard wires. Guard wires surrounding the three cells are included for a proper definition of the electric field inside the cells. 
The prototype is 20 cm long: four stainless-steel rods keep PCBs in position and stretch the wires.

The prototype, shown in figure~\ref{tritubonobox}, is inserted in a plexiglas gas-tight box equipped with two thin (1~mm) windows. The inner faces of the box are covered by a thin aluminium foil for electromagnetic and radiofrequency shielding of the wires.
The chamber prototype is operated with 
helium-isobutane 85--15 provided by pre-mixed gas tanks. 
The output signals are read out at one end of the sense wires via a 
prototype pre-amplifier of the final drift chamber~\cite{bib:preamp},
with a bandwidth of $\approx 700 \mathrm{\,MHz}$ and a gain of $\sim7$.
The opposite end is terminated with a 330 $\Omega$ resistor, to match the chamber characteristic impedance.

For the indirect measurement of the single hit resolution the pre-amplified signals are acquired through the Domino Ring Sampler (DRS) evaluation board~\cite{bib:drs4,bib:drs4evboard},
a waveform digitizer with a bandwidth of 700~MHz and a sampling speed of 2~Giga-samples per second (GSPS).
Wire voltages are set at 1700~V on the anode wires and 375~V on the guard wires, with respect to the grounded field wires, resulting in a gain in excess of several times $10^{5}$.
\begin{figure}[!t]
\centering
\includegraphics[width=0.5\textwidth]{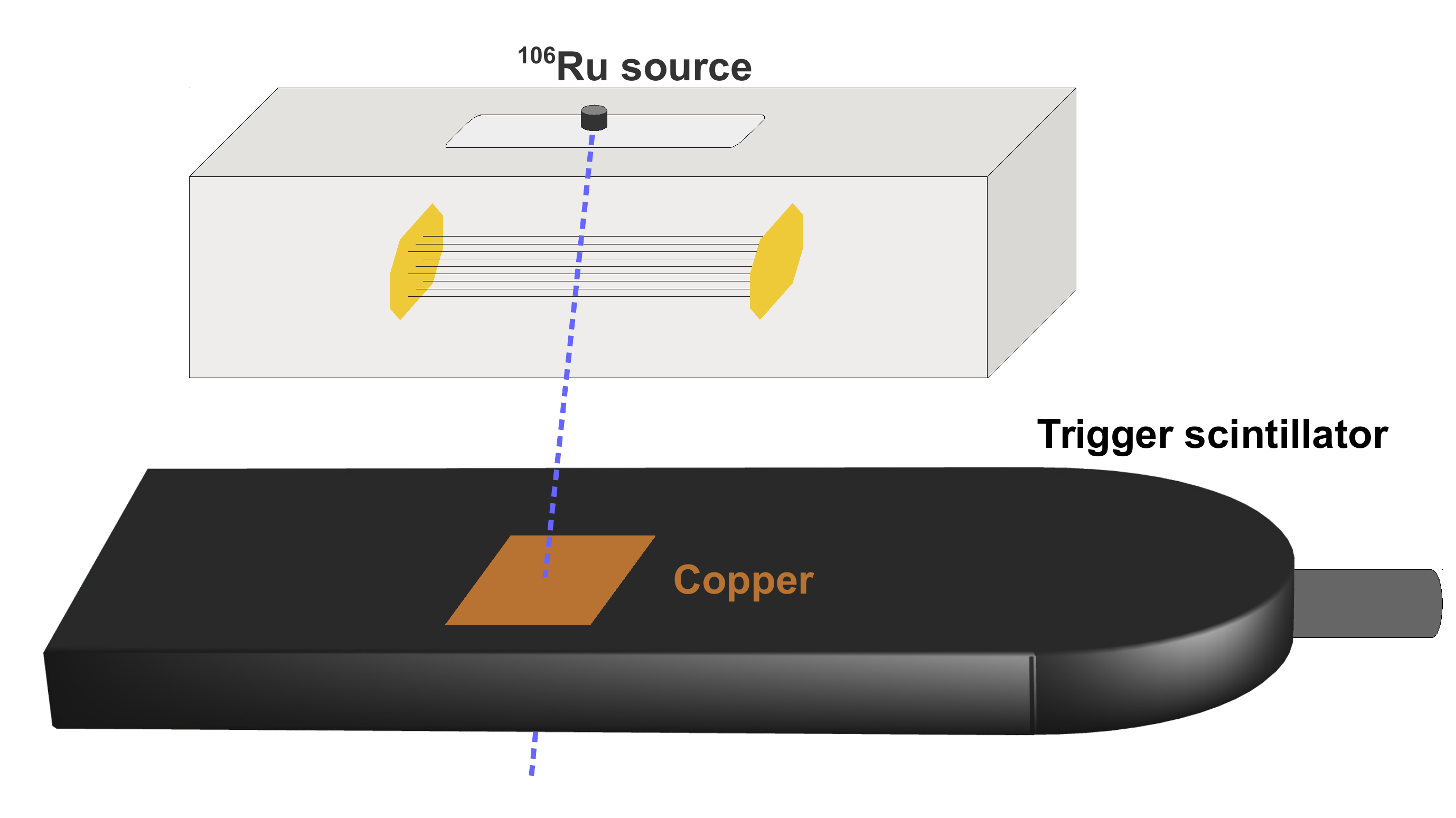}
\caption{Three-cells prototype. Schematic view of the experimental set-up (not in scale).}
\label{fig:tricellasetup}
\end{figure}
Measurements of time-to-distance relations and spatial resolution are performed by irradiating the prototype with a ruthenium source. A plastic scintillating counter is placed on the opposite side with respect to the radioactive source to provide a trigger signal (see figure~\ref{fig:tricellasetup}).
 The high-energy component of the $^{106}$Ru decay chain spectrum ($3\div 3.5$ MeV electrons from $^{106}$Rh $\beta$ decay) is selected by interposing a 500-$\mu$m copper foil between the prototype and the trigger scintillator. 
%%%%%%%%%%%%%%%%%%%%%%%%%%%%%%%%%%%%%%%%%%%%%%%%%%%%5
\subsection{Multi-cells prototype}
\begin{figure}[!b]
  \centering
 \subfigure[]{ \includegraphics[height=.3\textwidth]{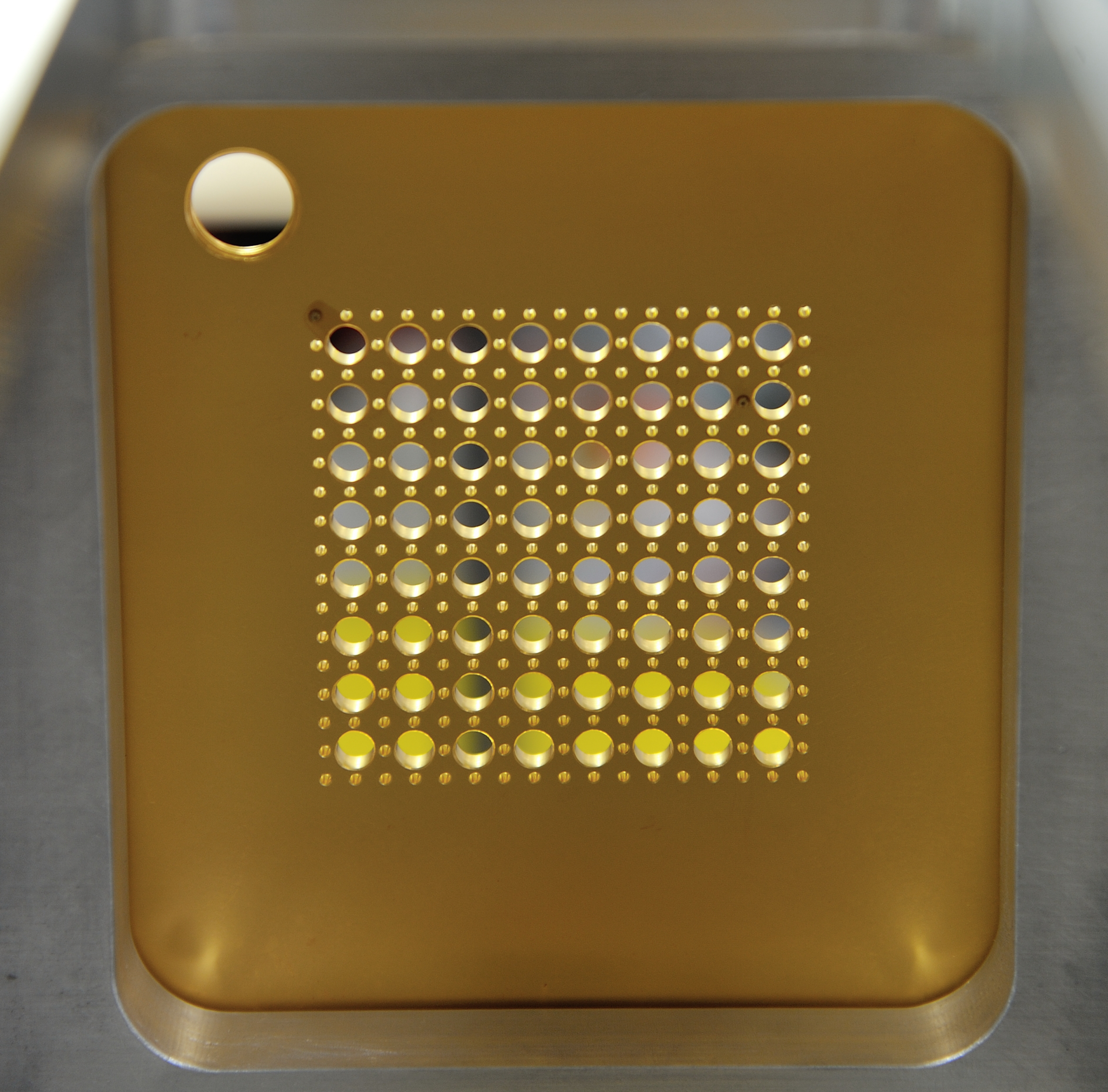}\label{fig:pictureProtoa} }
\hspace{2.5cm}
 \subfigure[]{ \includegraphics[height=.3\textwidth]{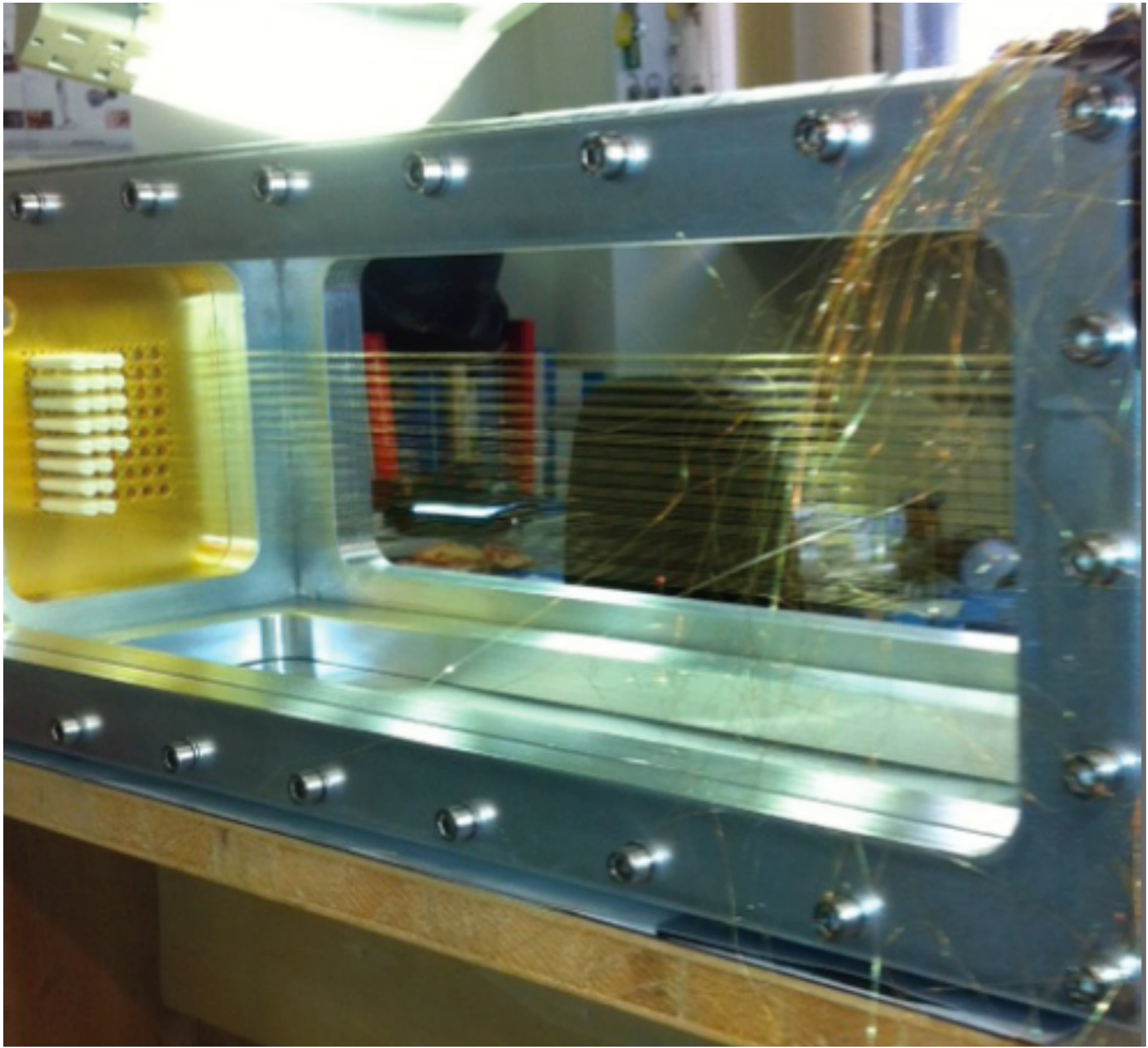}\label{fig:pictureProtob}}  
\caption{Multi-cells prototype. (a) Gold plated Al end-plate. (b) Internal view during the wiring procedure. }
 \label{fig:pictureProto}
  \end{figure}
  A third prototype is constituted by a gas-tight aluminium body of 200~$\times$~200~$\times$~500~mm$^3$. The lateral faces are made of 1.5~mm thick 
aluminium plates. The other two ends are closed by end-plates made of gold-plated aluminium, to facilitate ground contacts, that host the holes for gas inlet and outlet and the wire pins where the wires are soldered. The hole mask defines a 8 $\times$ 8 array of 7-mm side square cells, each with a sense wire 
(25~$\mu$m gold-plated tungsten) surrounded by 8 field wires (80~$\mu$m gold-plated tungsten). 
A view of one of the end-caps and an internal view during wiring are shown in figure~\ref{fig:pictureProto}.

The helium-isobutane mixture was set by two mass flow meters and, after calibration, was found to be 89--11.
We discuss in section~\ref{sec:resomix} the expected difference in resolution with respect to the nominal mixture.
The end of the sense wires at the high voltage side is terminated with the cell impedance (330 Ohm); the field wires are grounded, in electrical contact with the end-plates.
\begin{figure}[!t]
\centering
\includegraphics[width=120mm]{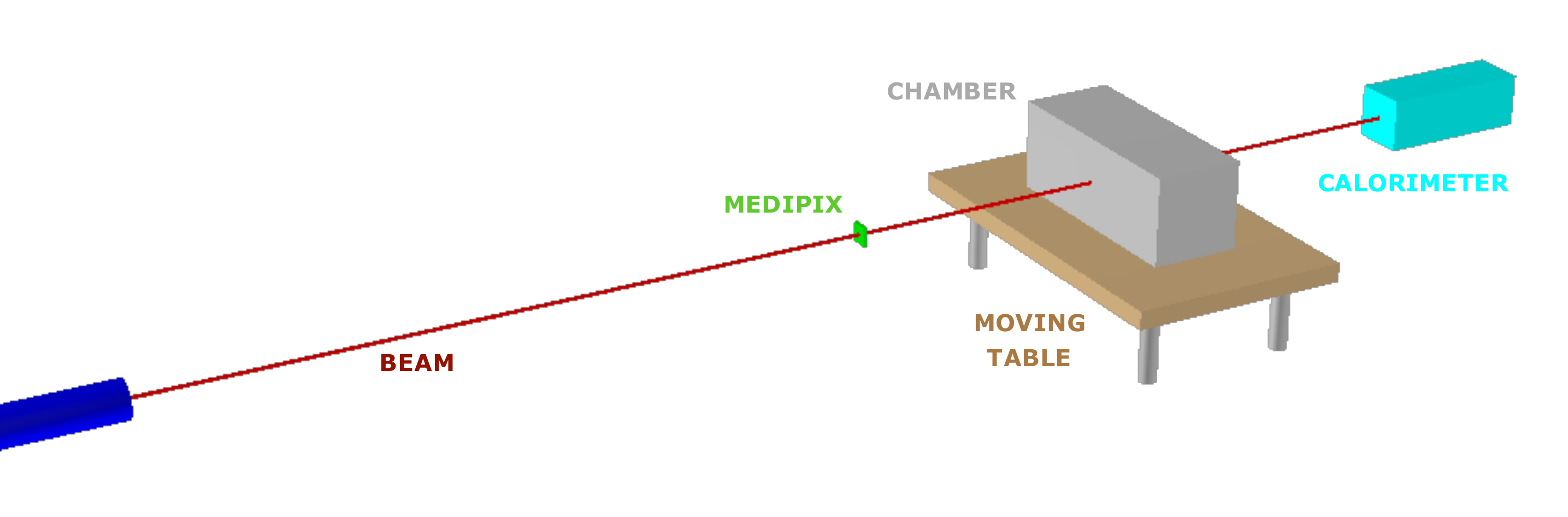}
\caption{Multi-cells prototype. Schematic view of the experimental set-up (not in scale).}
\label{fig:btf_setup}
\end{figure}
The operating voltage is set to 1620~V for a gain of about few $10^5$.
The front-end electronics is the same custom high bandwidth pre-amplifier described above;
the pre-amplified signals were read out by a DRS MEG board, with a 1~GHz bandwidth~\cite{bib:megdet}.
The multi-cells prototype was illuminated at the Beam Test Facility of the INFN Frascati National Laboratories~\cite{bib:btf}, 
where an electron beam of 447~MeV/$c$ momentum with an average multiplicity of one particle per spill was used.
The drift chamber prototype was placed perpendicularly to the beam axis, on
a precision moving table.
Upstream of the prototype a pixelated detector allowed the visualization of the beam spot while about 4~m downstream of the prototype a calorimeter was used to count %(and to set) 
the number of electrons per spill. A schematic view of the experimental set-up is shown in figure~\ref{fig:btf_setup}.
The beam rate was 25~Hz, driven by the LNF LINAC.
We chose a configuration with a beam spot as thin as possible in the vertical direction, in order to illuminate only a fraction of a cell. 

\section{Indirect estimates of single-hit resolution}
\label{sec:indirect}
%%%%%%%%%%%%%%%%%%%%%%%%%%%%%%%%%%%%%%%%%%%%%%%%%%%%%%%%%%

In all three prototypes the determination of the drift time of the first cluster is obtained with a double threshold algorithm: a low threshold is set at two or three times the RMS noise level, while a high threshold is set at five to seven times the same value (the particular number depending on the specific data set). The signal time is computed as the latest time at which the signal crosses the low threshold before the first crossing of the high threshold.
This procedure makes the measurement less sensitive to time-walk effects,
which increase with increasing threshold.
\begin{figure}[!t]
  \centering
  \subfigure[]{ \includegraphics[width=.47\textwidth]{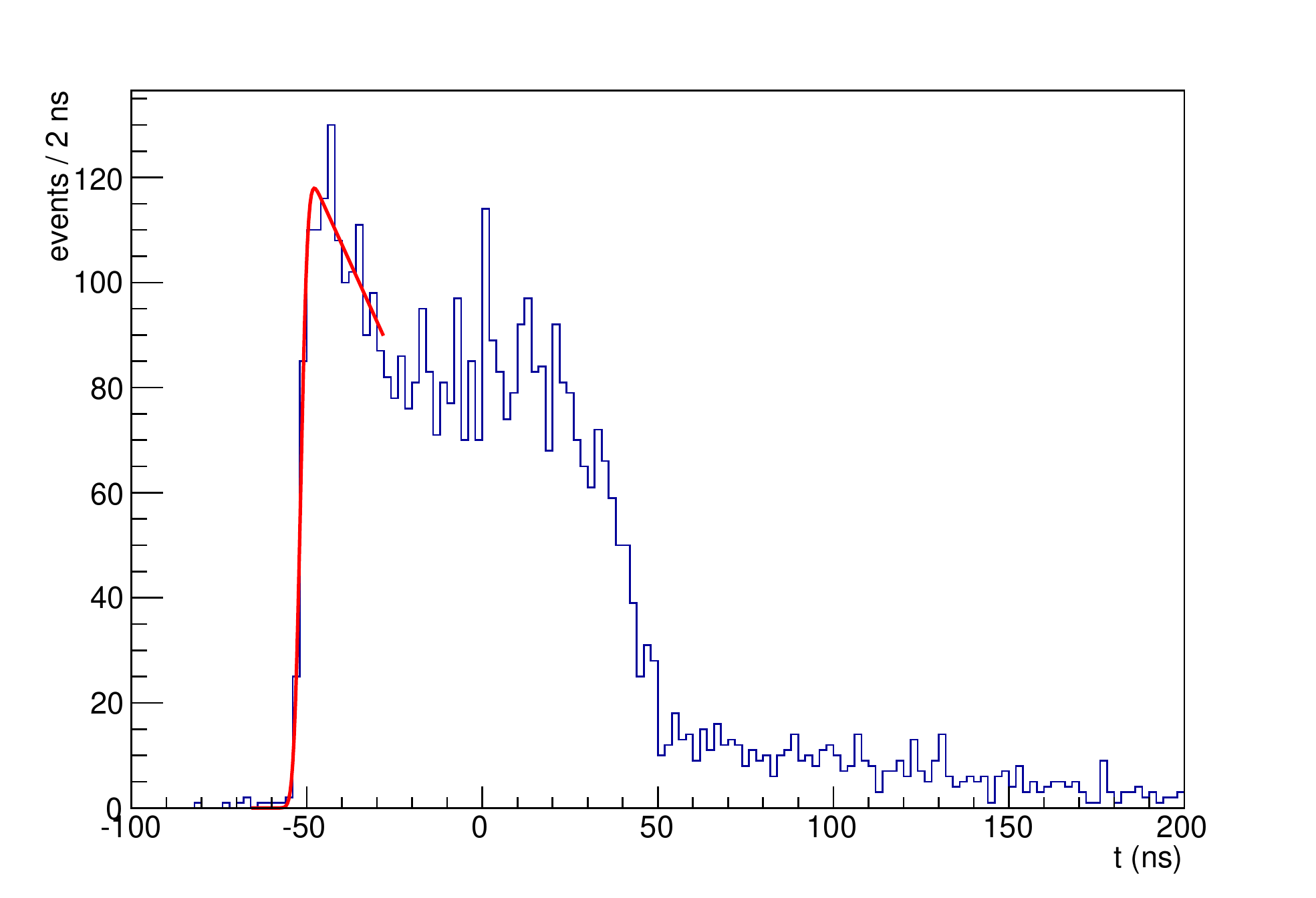} \label{mdtimes1750} }
  \subfigure[]{ \includegraphics[width=.47\textwidth]{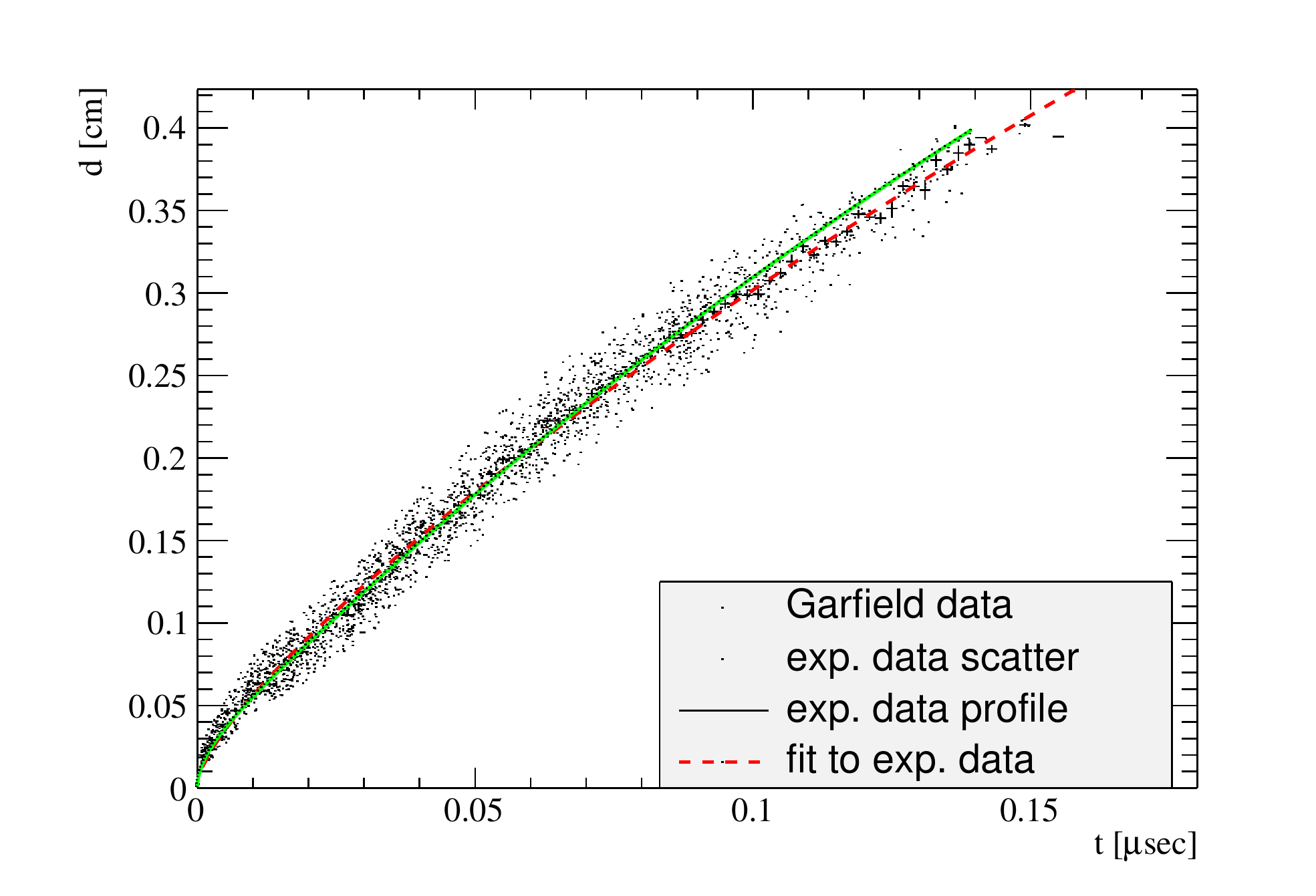} \label{xtttb1500}}
  \caption{(a) Distribution of the time of the signal subtracted by the trigger time, for a channel of the three-cells prototype. The red curve is the fit function described in the text. (b) Measured (dashed red) and simulated (solid green) drift time-space relation obtained with an iterative method for a channel of the three-tubes prototype.}
 \end{figure}
For each channel, the distribution of arrival times with respect to the trigger (see figure~\ref{mdtimes1750} for an example of the signal time distribution for one channel) is used to extract the relative delays ($t_0$), by fitting the edge of the drift time distribution with the product of an error function and of a first order polynomial
\begin{equation}
f_0 (t) = \mathrm{Freq}\left[\frac{t-t_0}{\sigma_0}\right]\times\left[p_0+p_1(t-t_0)\right].
\end{equation}
The parameter $t_0$ is the value obtained for zero drift time and $\sigma_0$ is the intrinsic drift time resolution (a few ns for the different prototypes).

An iterative procedure was used to determine the drift time-to-distance relations. Ionization electrons do not drift at constant speed due to the varying electric field, especially close to the sense wire. Therefore the time-to-distance relation does not follow a simple linear behaviour, being steeper at small drift distance, and it can be parametrized as a sum of powers of $t^{1/2}$:

\begin{equation}
 f_\mathrm{drift}(t) = a + b \sqrt{t} + c t +d \sqrt[3]{t}.
 \label{eq:rdit}
\end{equation}

The initial estimate of the time-to-distance function parameters is done by fitting the function~(\ref{eq:rdit}) to simulated data, made with Garfield (or Garfield++ relatively to each single prototype under study)~\cite{bib:garfield,bib:garfieldpp}, 
see figure~\ref{xtttb1500}. For the case of the three-tubes or the three-cells, using the nominal time-to-distance the drift distance in the central cell is measured averaging the drift distance of the two external cells and corrected for the nominal staggering value. The drift time of the central cell is put in relation with the drift distance, measured as described, to fit a new time-to-distance relation. The procedure is iterated until a convergence criterion is reached; in figure~\ref{xtttb1500} an example of the time-to-distance relation obtained for the three-tubes prototype is shown. Similar strategies are used for the other prototypes.

\subsection{Resolution measurements from the three-tubes and the three-cells array prototypes }\label{sec:tritube}
%%%%%%%%%%%%%%%%%%%%%%%%%%% LE %%%%%%%%%%%%%%%%%%%%%%%%%%%
\label{sec:resomix}
The three-tubes and three-cells array prototypes allow to measure the average (over almost all the drift distance range) single cell resolution. A preliminary measurement with the three-cells array was reported in~\cite{bib:oldtricell}. As shown in figure~\ref{fig:Tritube} for almost vertical tracks traversing all three cells and that do not pass in between the sense wires, the following relation is valid:
\begin{equation}
\mathcal{S}\equiv\frac{d_1+d_3}{2}-d_2 \simeq \pm \Delta
\label{triteq}
\end{equation}
where $\Delta$ is the stagger of the central wire (500 $\mu$m in our case) and $d_1, d_2, d_3$ are the impact parameters respectively on the first, second and third cell. Corrections due to non-verticality are below the percent level. 
\begin{figure}[!t]
  \centering
\subfigure[]{ \includegraphics[width=.47\textwidth]{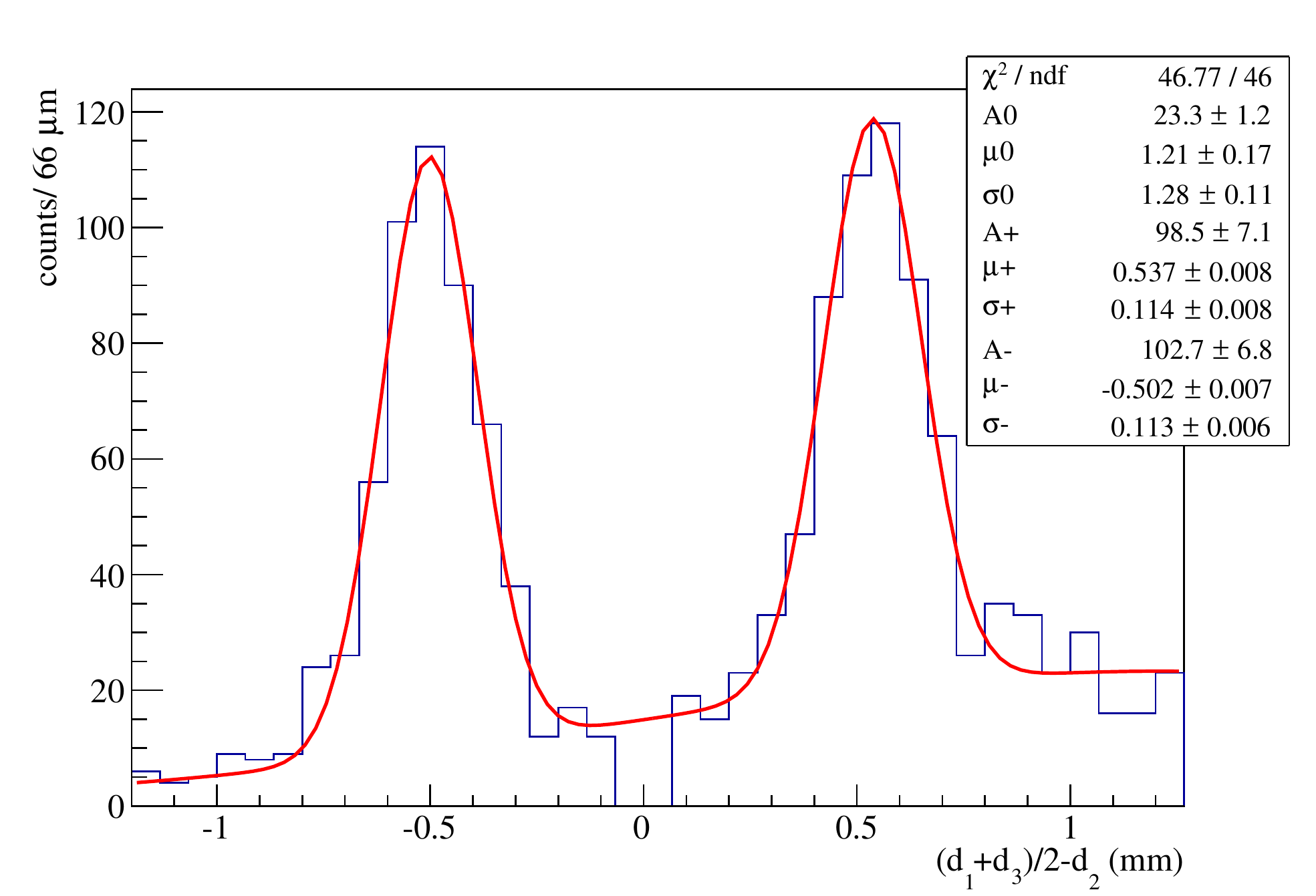} \label{bicorno-le} }
\subfigure[]{ \includegraphics[width=.47\textwidth]{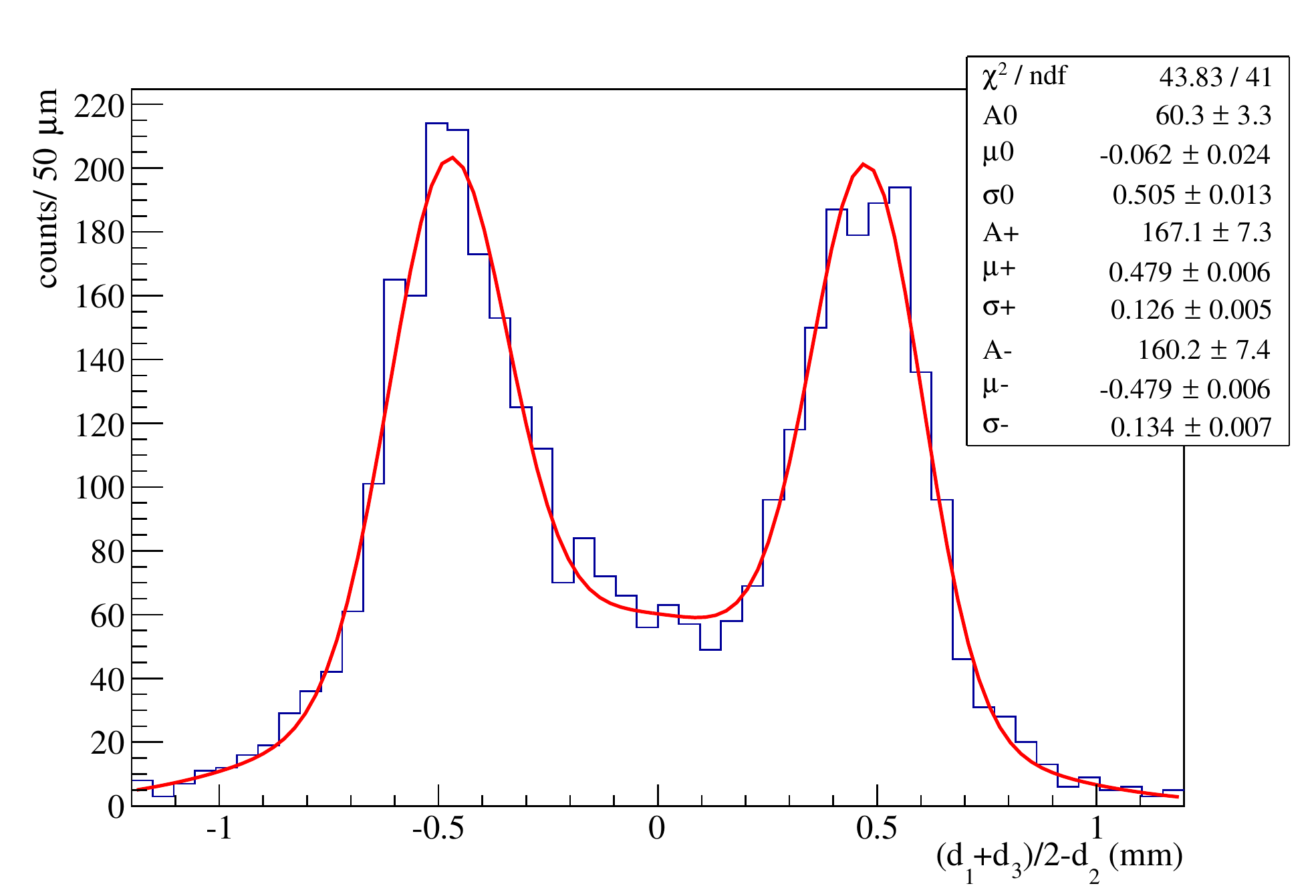} \label{bicorno-pi} }
  \caption{Three-tubes (a) and three-cells (b) measurement distribution with helium-isobutane 85\%--15\% superimposed with a 3-Gaussian fit.} \label{bicorno}

 \end{figure}
 The distribution of $\mathcal{S}$, shown in figure~\ref{bicorno-le} for the three-tubes prototype and in figure~\ref{bicorno-pi} for the three-cells prototype, is bimodal as expected from equation~(\ref{triteq}). It is well described by the sum of three Gaussian functions, two with means $\mu^{\pm} = \pm \Delta$ and one which takes into account wrong track reconstructions or tracks for which equation~(\ref{triteq}) does not hold, in particular those which are crossing the zone between the sense wires.
From the fitted value of the peaks position it is possible to measure the stagger value, approximately $520~\mu$m for the three-tubes and $480~\mu$m for the three-cells, in agreement with the expected stagger (500~$\mu$m) taking into account the uncertainties
 on the positioning of the anode wires.
Assuming an equal average resolution for the three cells ($\sigma_{d1}=\sigma_{d2}=\sigma_{d3}=\sigma_d$),
we obtain from equation~(\ref{triteq}) by using the peak widths $\sigma_\mathcal{S}^+$ and  $\sigma_\mathcal{S}^-$
 \begin{equation}
  \begin{split}
\sigma_d \simeq &~\sqrt{\frac{2}{3}}\,\left(\frac{\sigma^+_\mathcal{S}+\sigma^-_\mathcal{S}}{2}\right)\\
= &~93 \pm 5 \,\mu m\text{ \,for the three-tubes prototype}\\
= &~106 \pm 4 \,\mu m\text{ \,for the three-cells prototype}\\
\end{split}\label{resd}
 \end{equation}
The two measurements differ in both the geometry of the detector and the ionising particles used.
Since multiple Coulomb scattering (MS) strongly depends on both aspects, we evaluated the MS contribution to $\mathcal{S}$ with a GEANT4 simulation~\cite{bib:geant4}.
In the three-tubes, cosmic muons undergo MS in the copper walls while  for the MeV-electrons from the Ruthenium source MS occurs in the gas itself. In both cases the contribution to the resolution $\sigma_d$ is about 10~$\mu$m.
Since this quantity adds up in quadrature to the single hit resolution, the resulting MS contribution is below the uncertainty of the measurement. 
Concerning the different geometries of the detectors, two features are relevant for the average resolution measured. First, the three-tubes measurement is averaged over an interval in impact parameter wider than the one of the three-cells. Given that the resolution is higher at larger impact parameters, one can expect a better resolution for the three-tubes case.
Second, relation~(\ref{triteq}) does not hold for tracks passing in the proximity of the wires, where the resolution worsens, thus our average is expected to be better than the correct average single cell resolution. The size of the effect is determined comparing the result with that obtained using an external tracker.

\begin{table}[!t]

\caption{Measured spatial resolutions with different gas mixtures.}
\vspace{5pt}
\label{tabbella}
\centering
\begin{tabular}{|c|c|}%
\hline
Mixture  &  {$\sigma_{d}$ ($\mu$m)} \\% 
\hline 
50--50 & 71   $\pm$  2 \\
75--25 & 80  $\pm$  4 \\
80--20 & 90  $\pm$  4 \\
85--15 & 93  $\pm$  4 \\
90--10 & 107  $\pm$  7 \\
95--5  & 115  $\pm$  15 \\
\hline
\end{tabular}
\end{table}

The measurement with the three-tubes prototype was repeated with different helium/isobutane content, from 50--50 to 95--5, for studying the dependence of the single-hit resolution on the gas mixture. 
The resolutions obtained for the several gas mixtures are reported on Table~\ref{tabbella}.
We modelled the gas (ionisation) contribution to the single cells resolution as:
\begin{equation}
\sigma_d = \sigma_0 + \alpha/N
\end{equation}
where $N$ is the average number of ionisation clusters created in the gas volume, $\sigma_0$ and $\alpha$ are free parameters.  For a given helium fraction $f_\mathrm{He}$ of the mixture, we have $N=f_\mathrm{He}N_\mathrm{He}+N_\mathrm{Ib}(1-f_\mathrm{He})$, where $N_\mathrm{He}=7.4/$cm, $N_\mathrm{Ib}=54/$cm are the average cluster densities for helium and isobutane, respectively. 
\begin{figure}[!b]
\begin{centering}
\includegraphics[width=0.7\textwidth]{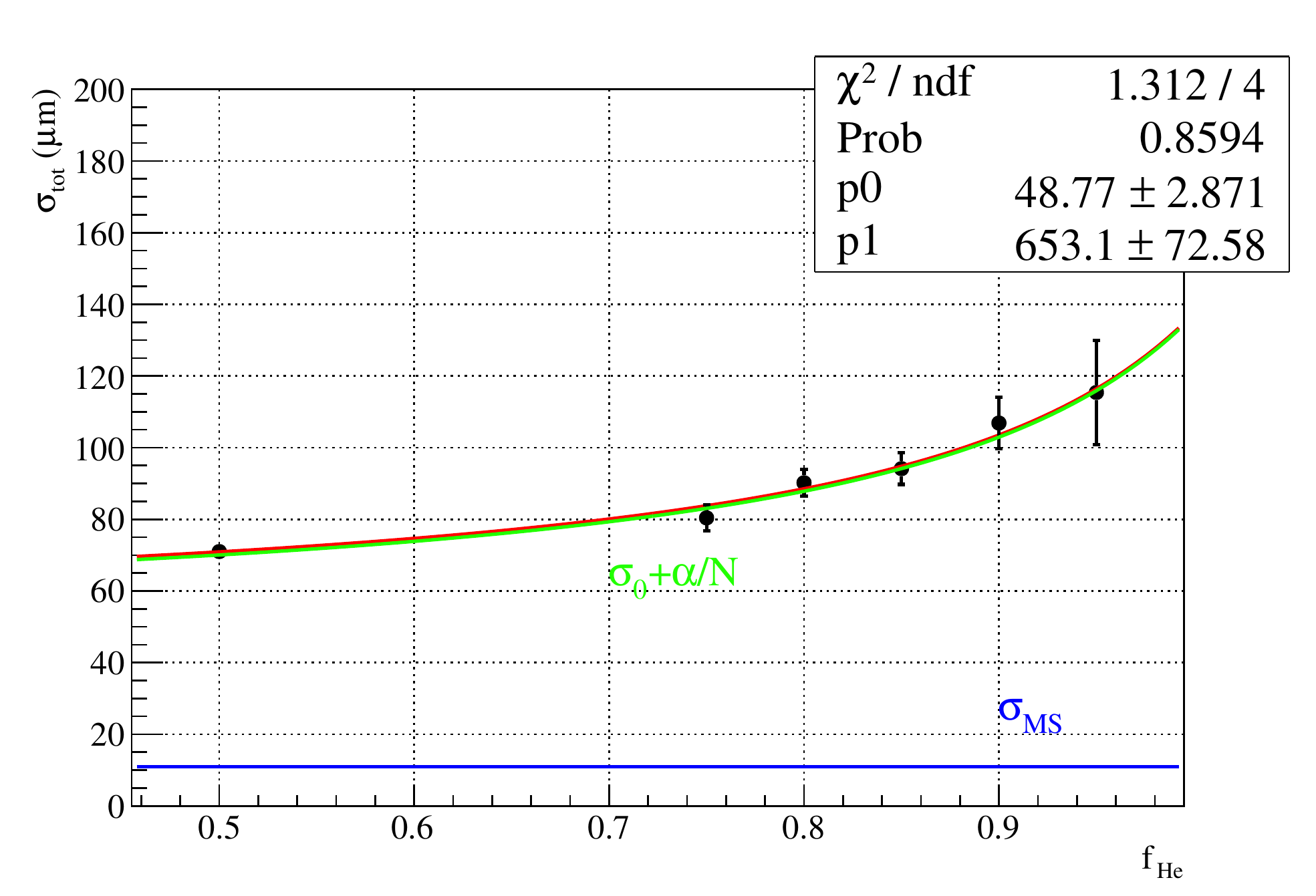}
\par\end{centering}
\protect\caption{\label{fig:Tritube-res} The resolution $\sigma_{i}$ as a function
of the helium content of the mixture.}
\end{figure}
The MS contribution due to the copper walls was added in quadrature:
\begin{equation}
\sigma_\mathrm{TOT} = \sqrt{\sigma_\mathrm{MS}^2 + \left(\sigma_0 + \alpha/N\right)^2}
\label{sigmatot}
\end{equation}
In figure~\ref{fig:Tritube-res} data are well described by function~(\ref{sigmatot}).
Therefore the expected value for the mixture 89--11 used in the multi-cells prototype is approximately 100~$\mu$m.

%%%%%%%%%%%%%%%%%%%%%%%%%%%%%%%%%%%%%%%%%%%%%%%%%%%%%%%%%%
\subsection{Resolution measurement from multiple-cells prototype}

For the multi-cells prototype data were taken with the instrumented cell rows parallel to the beam axis. In this configuration, each track went through 
one row only, with few exceptions. Since there is no staggering between cells in the same row, and an external tracker was not available, in this position it
is not possible to distinguish tracks passing above or below a given wire on an event-by-event basis. Hence, for calibration purposes some data were taken 
with the chamber tilted at about $17^\circ$ with respect to the beam axis in order to have tracks crossing different rows.
\begin{figure}[!t]
\centering
\subfigure[]{\includegraphics[width=56mm]{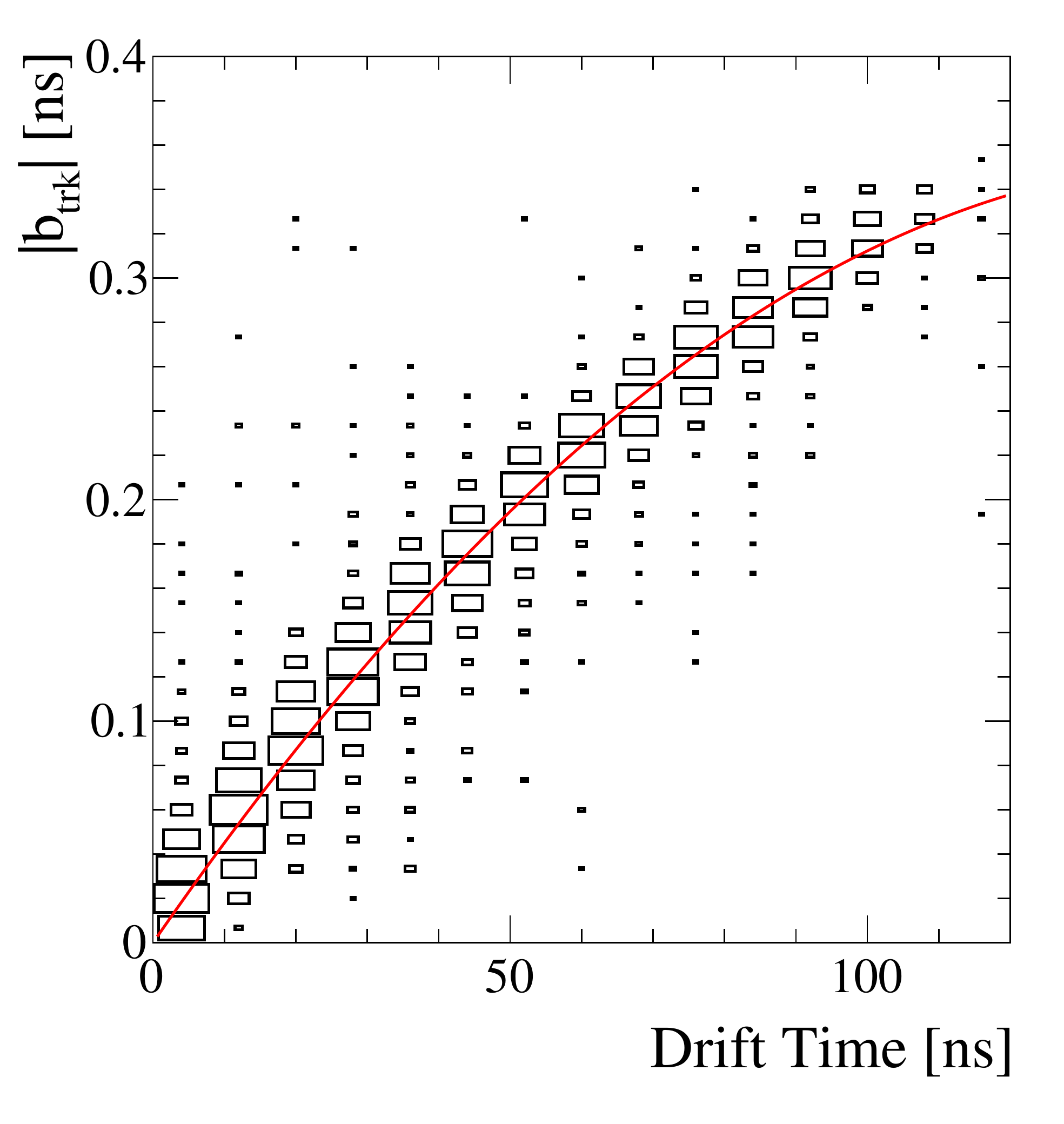}}
\subfigure[]{\includegraphics[width=60mm]{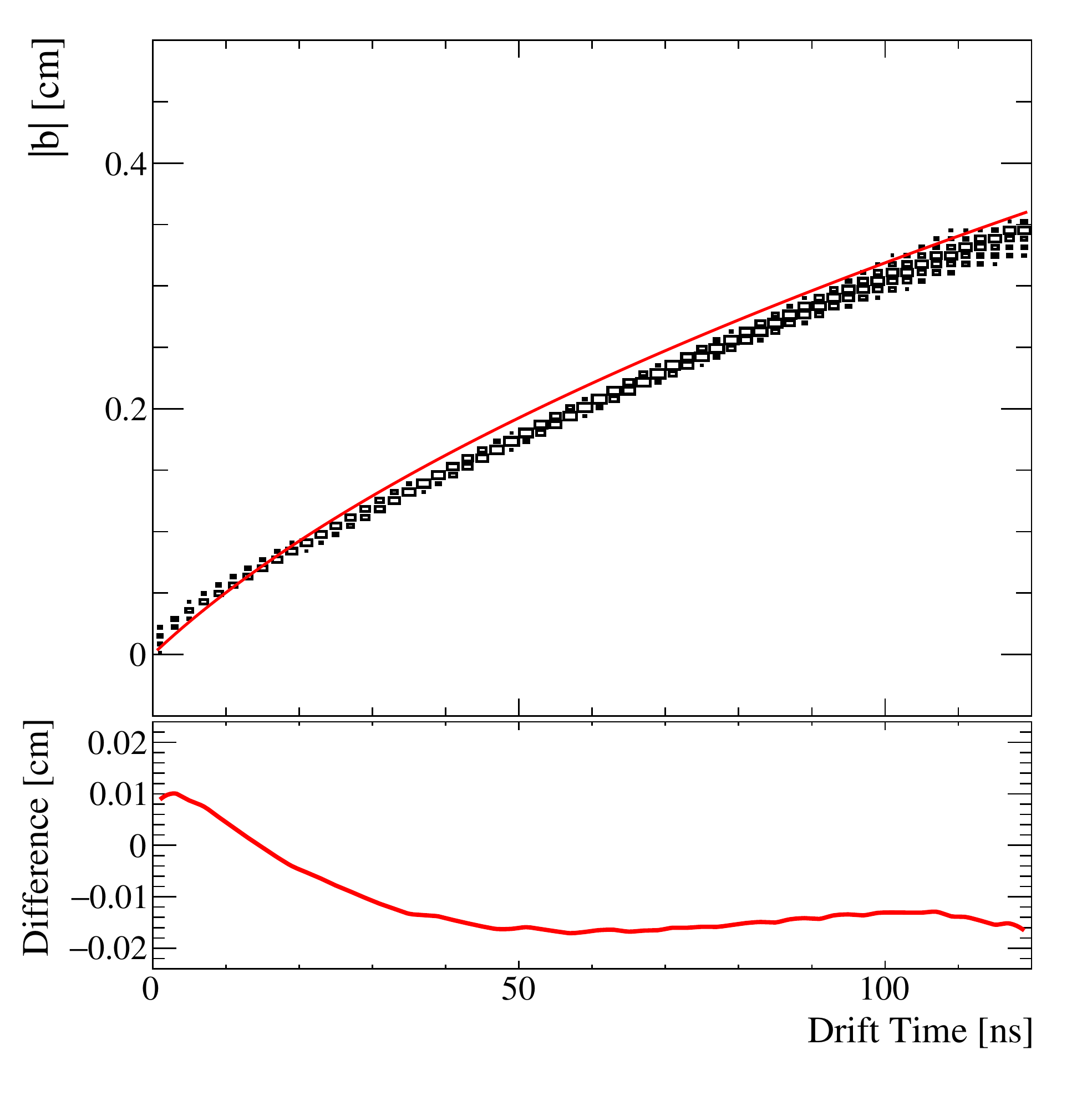}}
\caption{Multi-cells prototype. (a) Self estimate of the time-to-distance relation: the track estimate of the impact parameter versus the drift time (black) is fitted on data with a second order polynomial 
function (red). (b) Self estimate (red) and GARFIELD simulation (black) of the time-to-distance relation. 
}
\label{fig:txy}
\end{figure} 
\begin{figure}[!b]
\centering
\includegraphics[width=.45\textwidth]{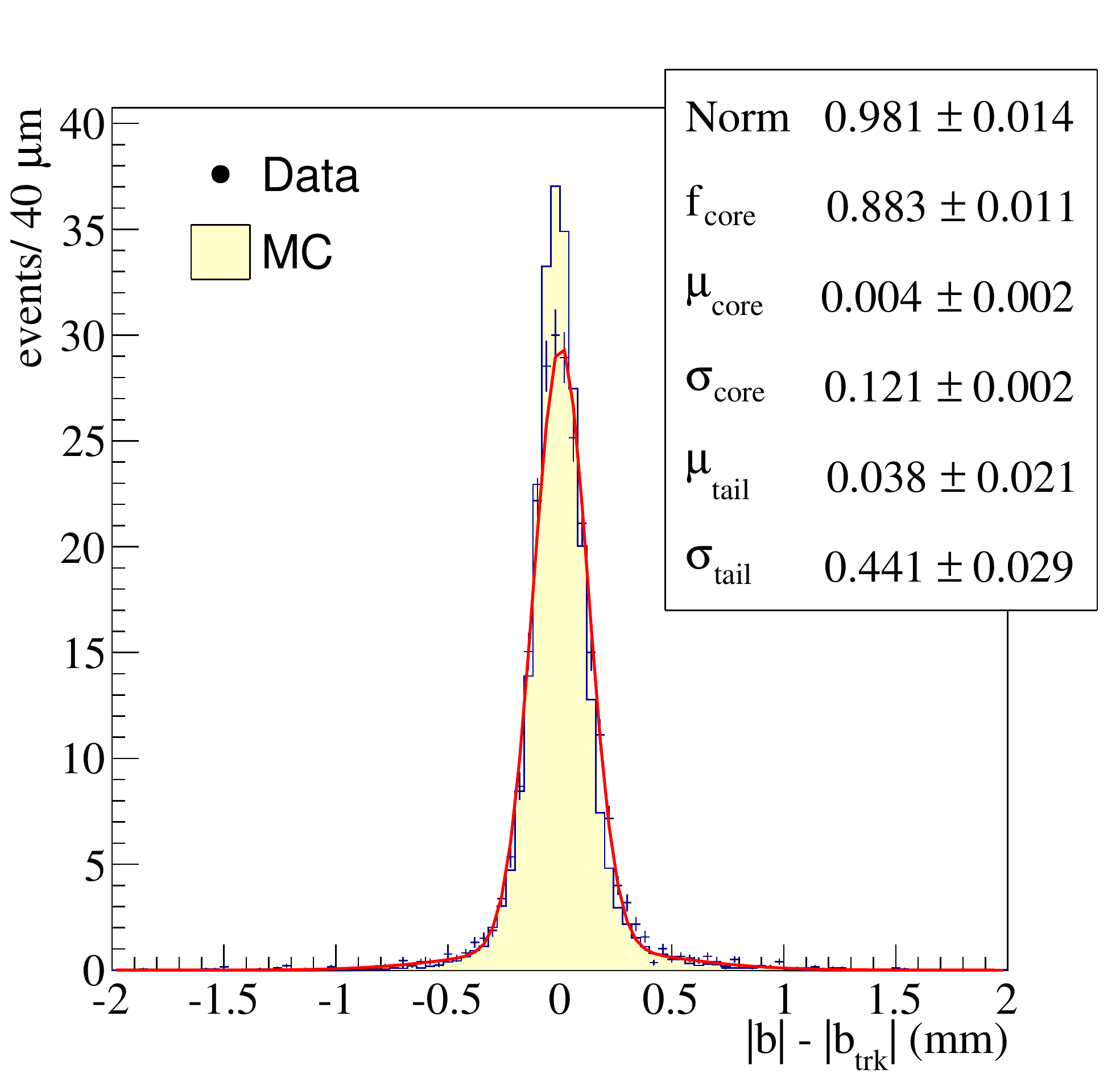}
\caption{Spatial resolution obtained with multi-cells prototype in horizontal position at BTF (black dots) fitted with a double Gaussian function (red line), compared with the resolution obtained on a Monte Carlo (MC) simulation (histogram).}
\label{fig:reso_all}
\end{figure}

Misalignments within the tolerances of the construction procedures ($\sim100~\mu$m) have been found and corrected
with an iterative alignment procedure. Figure ~\ref{fig:txy} shows the time-to-distance relation.
After calibration of the detector, the resolution can be determined by looking at the distribution of the hit-to-track residuals which is shown in figure~\ref{fig:reso_all} for the data taken with the prototype in horizontal position. A comparison with a simulation which reproduces the expected ionization pattern of 447~MeV/c electrons, the geometry of the prototype, the measured noise and signal amplitude, the measured misalignments and time-to-distance relation is super-imposed.
The measured resolution includes a non negligible contribution coming from the estimate of the drift distance from the other three hits in the track. % 
From a comparison of the measured and true resolution in MC, we expect the measured core resolution to be overestimated by about 25$\%$. The true core resolution 
of the detector therefore amounts to about 100~$\mu$m in our data.

\section{Single-hit resolution direct measurement}
\label{sec:direct}

\subsection{Cosmic Ray Telescope Facility}
\begin{figure}[!b]
  \centering
   \includegraphics[width=\textwidth]{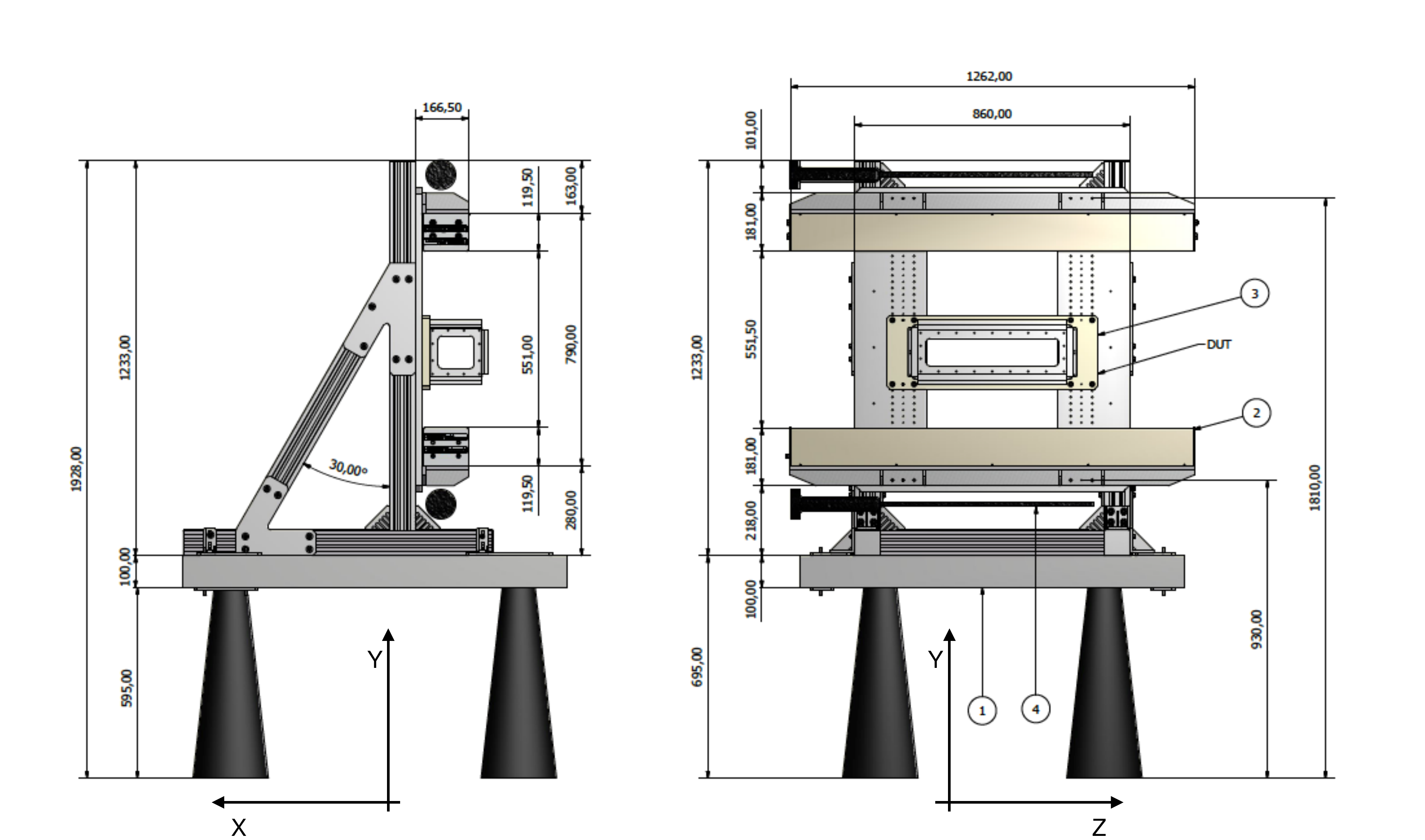} 
  \caption{$xy$ and $yz$ views of the cosmic ray telescope. The numbers indicate the granite table (1), the SVT modules (2), the DUT (3) and a scintillating slab (4).}
  \label{TeleImage}
 \end{figure}

\begin{figure}[!t]
  \centering
 \subfigure[]{  \includegraphics[height=.35\textwidth]{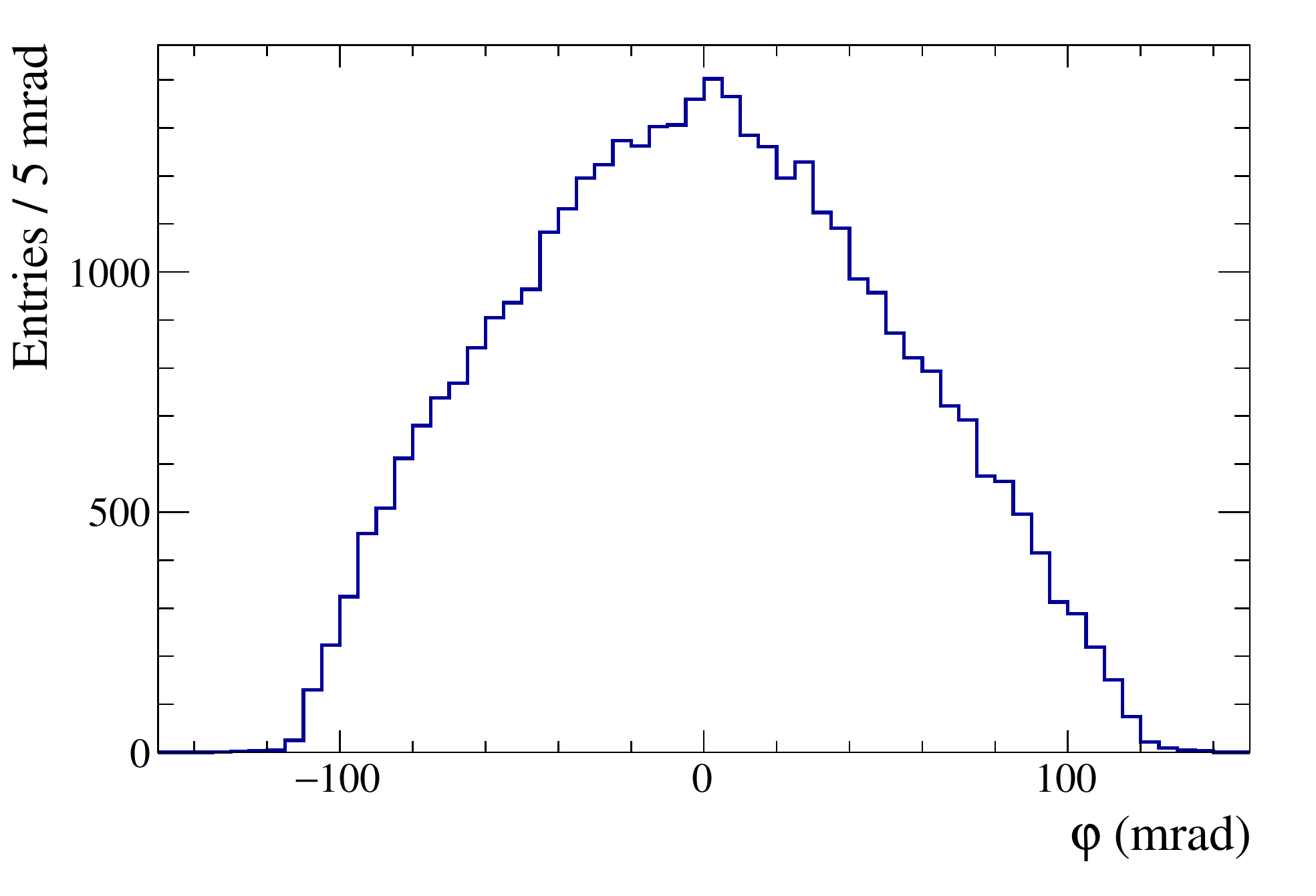} \label{phidistrib}  }
  \subfigure[]{ \includegraphics[height=.41\textwidth,angle=0]{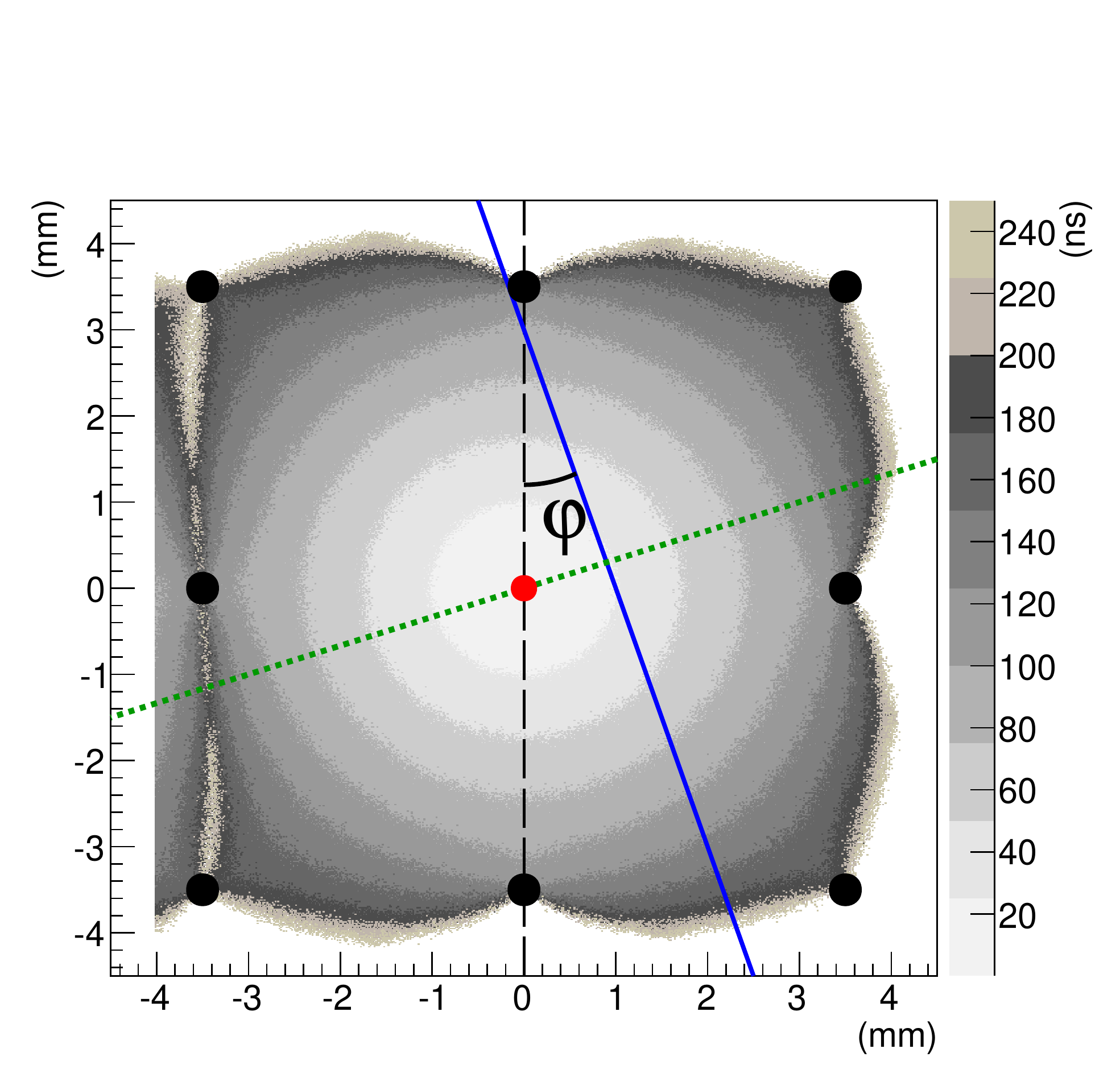}  \label{dtimesphi}}
   \caption{(a) Angular distribution of the tracks reconstructed by the telescope in the $xy$ view ($\varphi$ angle). (b) Drift time distribution for electrons in the cell, obtained with Garfield++. Tracks impinging at an angle $\varphi$ (solid blue line) have impact parameters lying on the dotted green line.}
 \end{figure}

A direct measurement of the single-hit resolution requires an additional tracking device able to provide an independent measurement of the impact parameter in the cell with a resolution better than that of the detector under test (DUT).
We used a cosmic ray tracker assembled at INFN Sezione di Pisa~\cite{bib:telescopio,bib:tricell}. 
The tracker consists of four planes
of double-sided silicon strip detectors with orthogonal strips.
 The detectors were originally built to serve as spare modules for the Silicon Vertex
Tracker of the BaBar experiment~\cite{bib:babar}.
The single-hit resolution of the modules is dominated by the strip pitches, being
100 and 210 $\mu$m in the two views (on the transverse plane and along the wire respectively), with  a high single-hit
efficiency varying from from 95\% to 99\%. In the cosmic ray tracker
configuration, the expected resolution of the track position
extrapolated to the DUT is roughly 24~and~35~$\mu$m in the two views, after taking into account effects
such as multiple scattering and internal alignment errors. For an expected single hit resolution of about $100~\mu m$, the MS effect is of the order of $3\%$. %\cite{bib:telescopio}.
Two scintillation counters, read by fast PMTs, are placed above and below the
system to provide both the trigger for cosmic muons and the
timing in the event reconstruction. The whole apparatus is visible in figure~\ref{TeleImage}, while
Figure \ref{phidistrib} shows the distribution of the $\varphi$ angle (defined in figure \ref{dtimesphi}) of the reconstructed 
tracks on the plane orthogonal to the wires.
 A computer receives both the track hits from telescope readout electronics and the differential signals from the three-cell
tracking device prototype, digitized by a V1729 VME board (2 GSPS, 300 MHz bandwidth). The signal of the PMT of the upper scintillating slab is digitized as well for a precise trigger timing.
For each run a file is produced with all the raw information needed for the offline analysis.

The same three-cells prototype used for the indirect measurement of the single-hit resolution discussed in section~\ref{sec:tricella} is placed in the telescope. 
A high voltage value of 1600 V (350 V) was set on the sense (guard) wires, corresponding to a gain of about few $10^5$, close to the design value of the full drift chamber. The measured rate on the single cell is about 1 track per minute.

\subsection{Timing and time calibration}

The determination of the drift time of the first cluster is obtained with the double threshold algorithm explained in the previous section.
As an additional improvement, a linear interpolation is performed between the sampled values returned by the digitizer in order to reduce quantization noise. 
Trigger time is obtained through a Gaussian fit of the PMT signal (the PMT signal shape is almost constant), which is then corrected for the time-of-flight of the cosmic ray and by the propagation time in the scintillator.
The time calibration is performed for each channel as explained in section~\ref{sec:indirect}. For all the channels an average value of $\sigma_0 \simeq 500$~ps is obtained. %

The external measurement of the impact parameter of cosmic rays on the cells requires the determination of the wire positions in the telescope Reference Frame (RF) $(xyz)$, defined in figure~\ref{TeleImage}. 
After an alignment procedure discussed in Appendix~\ref{app:align}, wire positions are determined with an uncertainty of $\sim 150\,\mu$m on the $x$ coordinate and of $\sim 300\,\mu$m on the $y$ coordinate.

\subsection{Measurement of time-to-distance relations}
The knowledge of the wire position allows us to calculate the time-to-distance relations  $f_\mathrm{drift}(t)$. In principle, due to the squared geometry, these curves depend on the impinging angle $\varphi$ (see figure~\ref{dtimesphi}). For perfectly squared drift cells the symmetry of the electric field (thus of the drift velocity) requires
\begin{equation}
f_\mathrm{drift}(t|\varphi) = f_\mathrm{drift}(t|-\varphi)=f_\mathrm{drift}\left(t\left|  \,^{\pi}\!/\!_{2}\right.-\varphi\right) 
\end{equation}

Garfield++ simulations are performed for evaluating the differences in drift times for electrons drifting  at $\varphi =0$ and those drifting at $\varphi =^\pi\!\!/\!_4$. According to simulations, the two drift curves overlap up to impact parameters of about 3 mm; for larger impact parameters the increase of the electric field close to the cathodes makes the drift faster for tracks at $\varphi =0^\circ$. % 
  \begin{figure}[!t]
  \centering
\subfigure[]{ \includegraphics[width=.47\textwidth,angle=0]{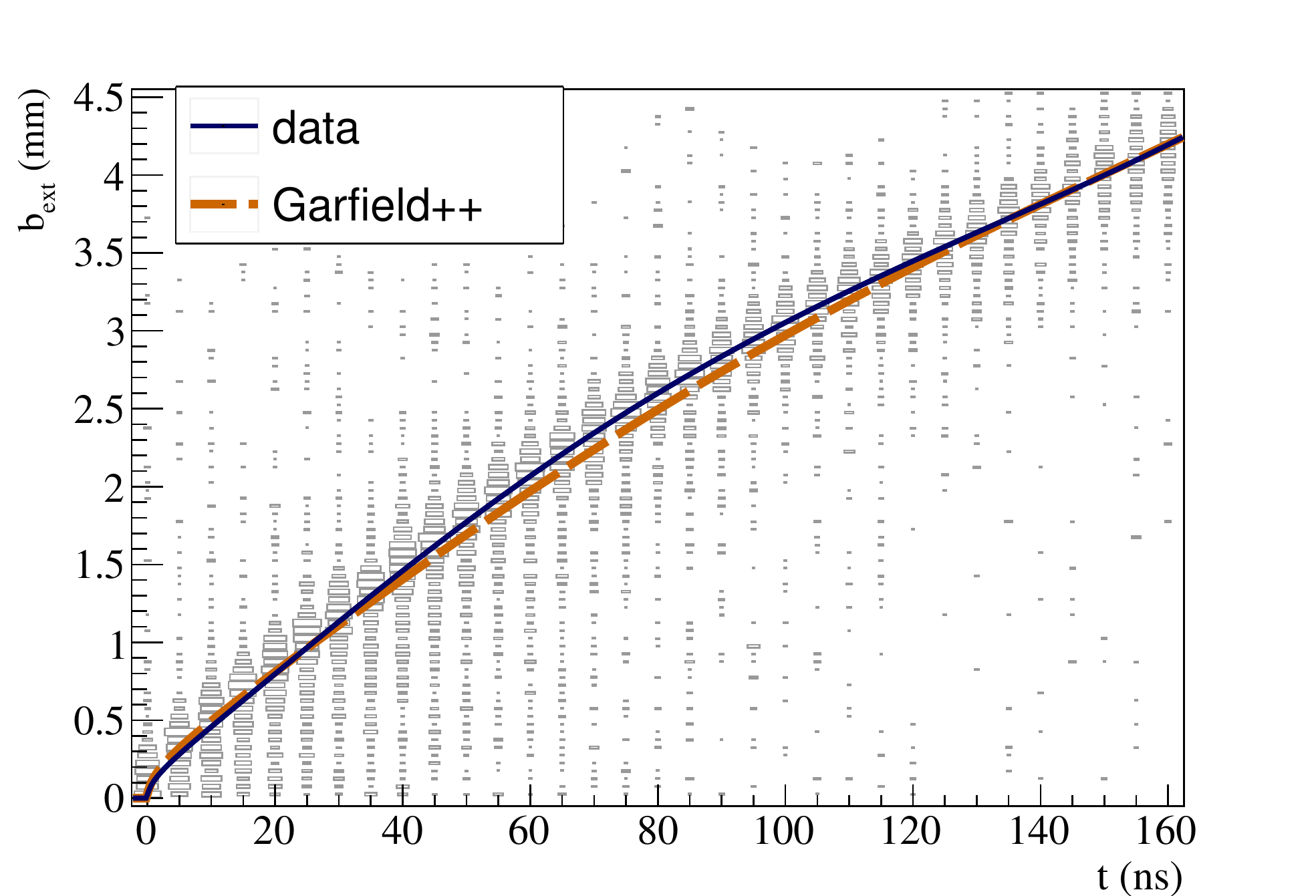} \label{DriftCurveObl}}
\subfigure[]{ \includegraphics[width=.47\textwidth,angle=0]{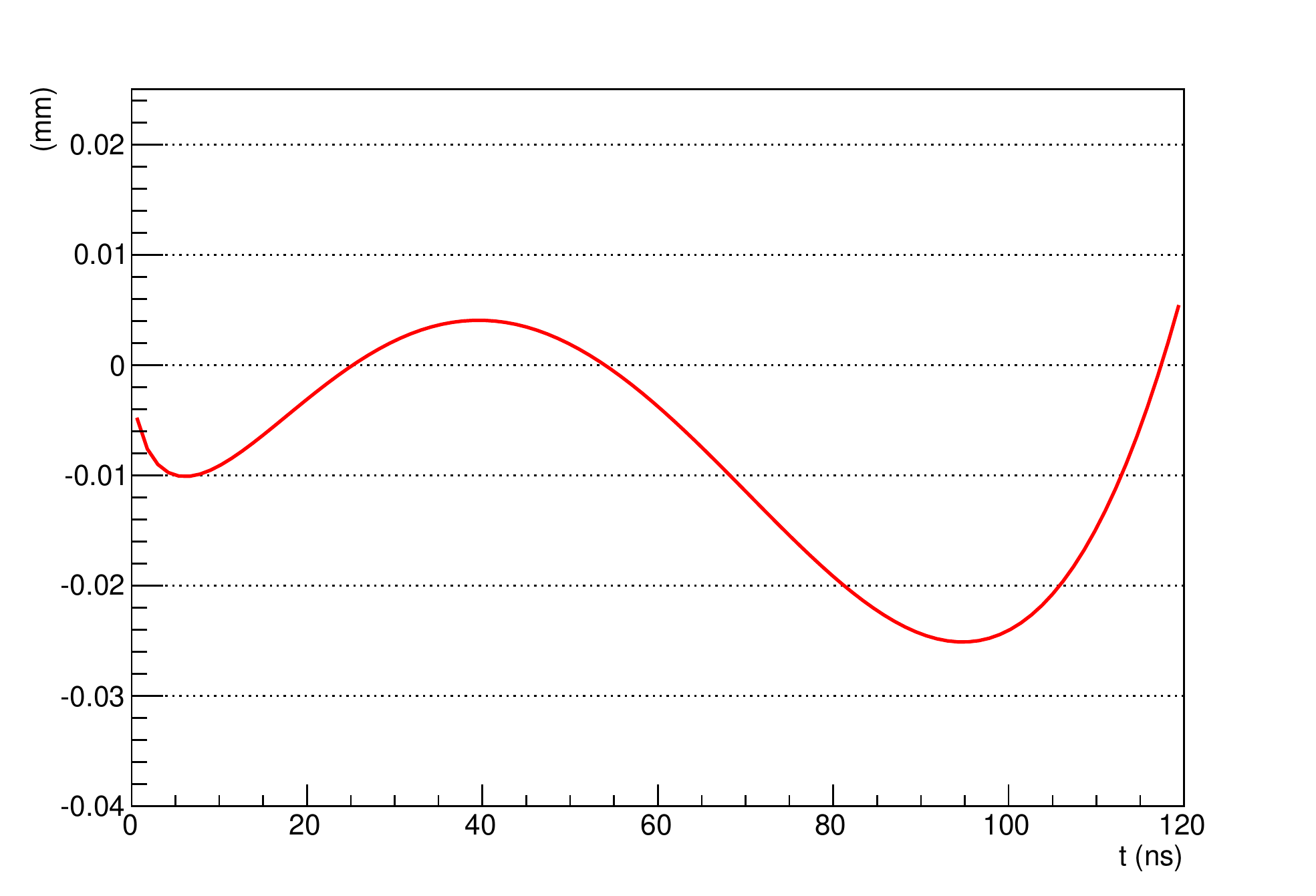} \label{ResDriftCurveObl}}
  \caption{(a) % 
  Simulated and measured drift curves for $\varphi=\pi/4$. (b) Residuals between measured drift curves for $\varphi=0$ and $\varphi=\pi/4$.}
 \end{figure}  
 Since the $\varphi$ angle spread of tracks in our set-up is about 200~mrad (as shown in figure~\ref{phidistrib}),
an investigation of the drift curve on the whole cell requires taking several runs with different orientations of the prototype. 
 In order to evaluate the dependence of the drift curve on the impinging angle
we performed two runs in the extreme cases of $\varphi = 0$ and $\varphi = \pi/4$.
 For determining an analytic curve fitting to data, we use  5-ns bins in drift time and perform a Gaussian fit to the impact parameter distribution in each bin. The fitted mean values are then plotted as a function of the bin centre value and fitted to an empirical function of the form 
\begin{equation}
f_\mathrm{drift} (t)= \sqrt{t}(p_0+p_1t+p_2t^2+p_3t^3)
\end{equation}
Noise events are cut out by setting a lower threshold on the waveform charge.
The selection of tracks that are well reconstructed by the telescope is performed through a cut on the sum of the values of the $\chi^2$ of the fits in the two views  of the telescope.
The box plot in figure~\ref{DriftCurveObl} shows that for small drift times the distribution of events cannot be well described by a Gaussian function since it is not symmetric: our technique of determining the curve introduces therefore a distortion for small impact parameters, taken into account by requiring the  function to be zero at $t = 0$.
The curves obtained in the two configurations ($\varphi = 0$ and $\varphi = \pi/4$) overlap to good accuracy, the largest difference being about 25 $\mu$m (see figure~\ref{ResDriftCurveObl}). 
Such a difference is quite smaller than the single-hit resolution, therefore we can use a $\varphi$-independent time-to-distance relation.
\subsection{Single-hit resolution measurement}
\begin{figure}[!t]
  \centering
   \includegraphics[width=.49\textwidth,angle=0]{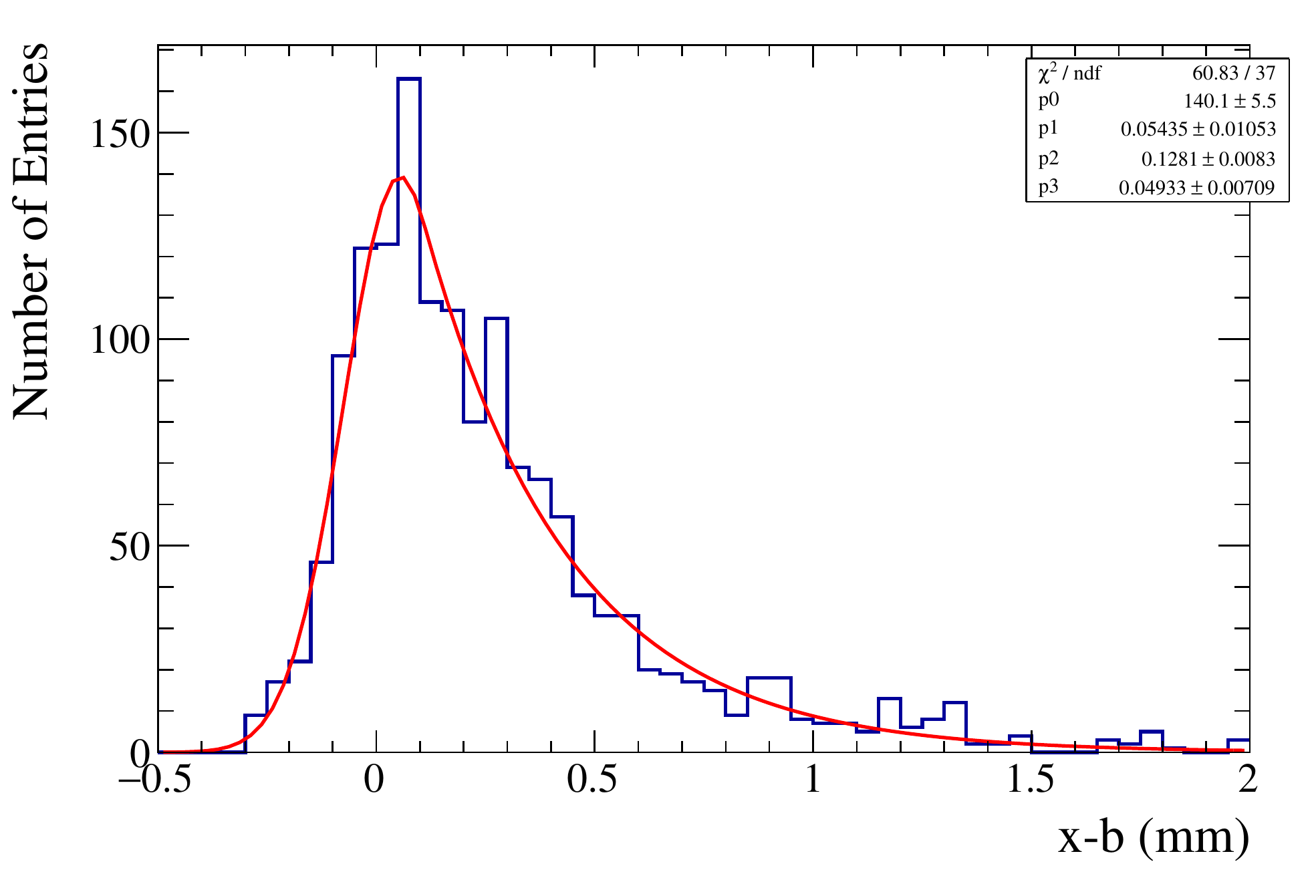} \,
         \includegraphics[width=.49\textwidth,angle=0]{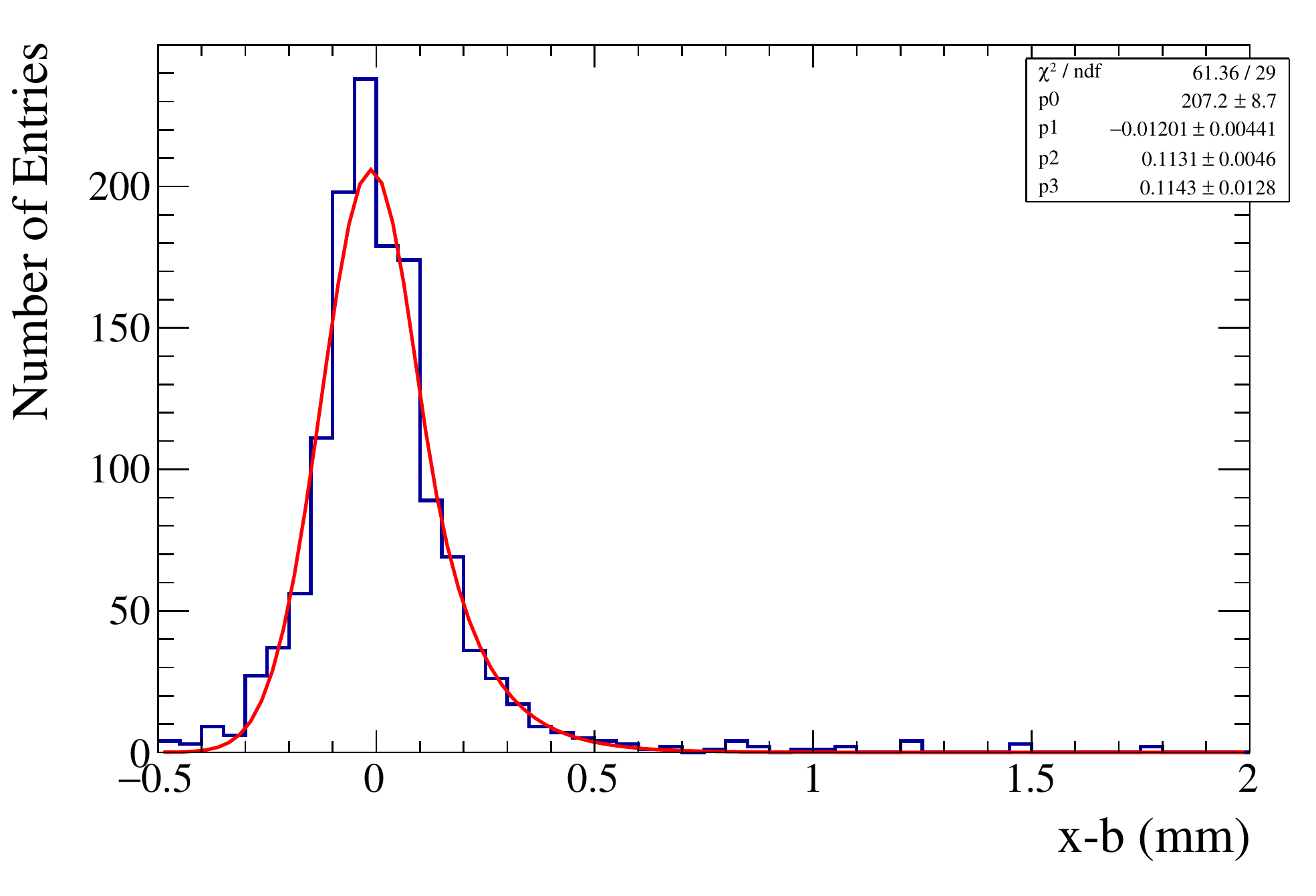}
  \caption{Distribution of the difference between the impact parameters returned by the drift chamber and the telescope in two intervals of impact parameter [0,~0.3] % 
  and [1.8,~2.1]~mm. Distributions are well described by the Gaussian/exponential function discussed in the text.}
  \label{nongaus}
 \end{figure}

 The largest contribution to the position resolution with very small cells filled with a low gas mixture
Since we have small cells filled with a low mass gas mixture, the largest contribution 
 comes from primary ionization fluctuations. This affects both the width and shape of the distribution of residuals on impact parameters: a bias is introduced in the estimate of the impact parameter, since drift distances are always larger than the true impact parameter. Due to geometrical effects, this behaviour results in a non-gaussianity of this distribution that increases at decreasing impact parameters, as shown in figure~\ref{nongaus}.
As a parametrization of the curve we used a function consisting of a Gaussian matched with an exponential at a distance $\delta$ from the Gaussian mean, which well describes data. The constraints of continuity and derivability in the matching point fix the parameters of the exponential function; the parameters of the function are therefore the two parameters of the Gaussian and $\delta$.
\begin{figure}[!b]
  \centering

\subfigure[]{   \includegraphics[height=.32\textwidth,angle=0]{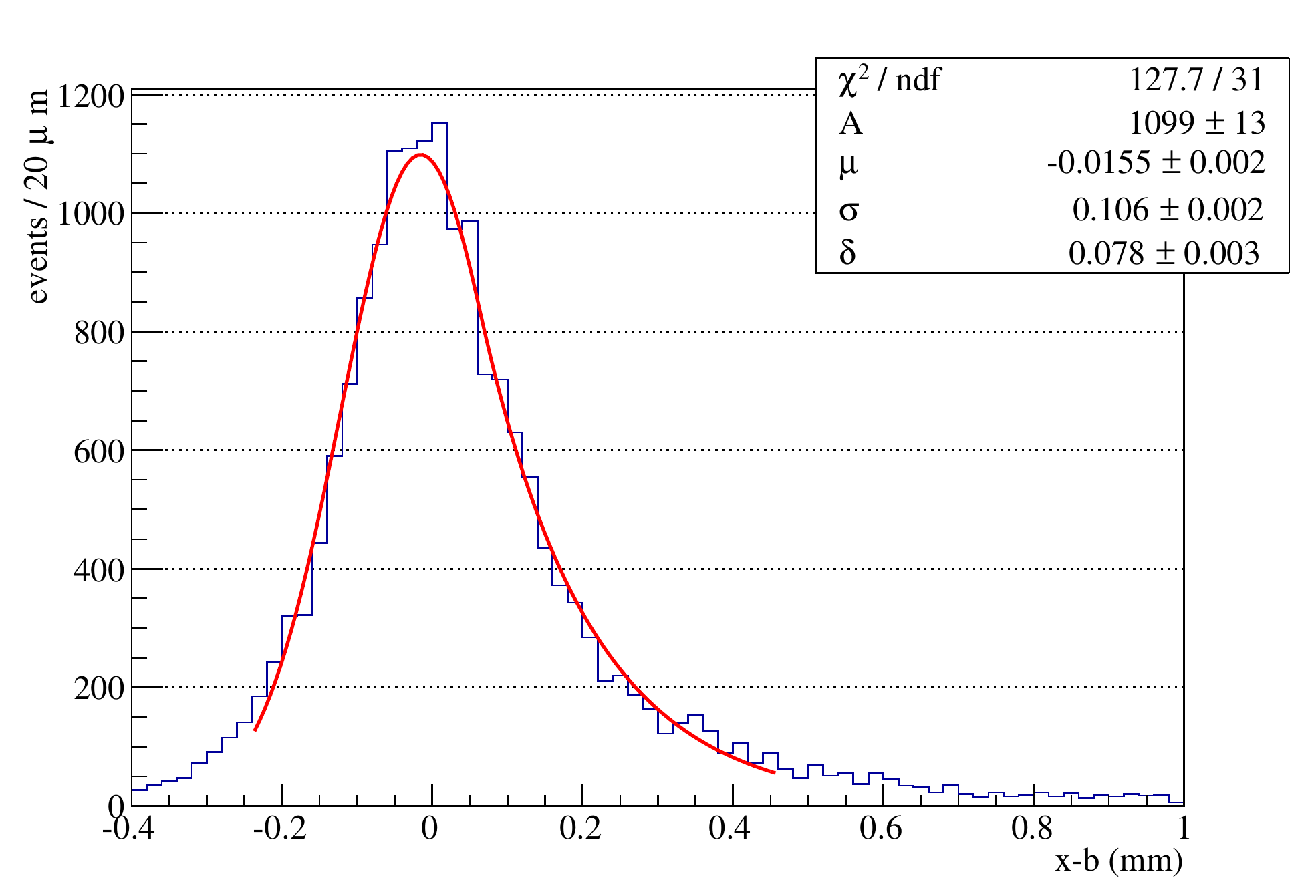}  \label{allreso}}
 \subfigure[]{   \includegraphics[height=.33\textwidth,angle=0]{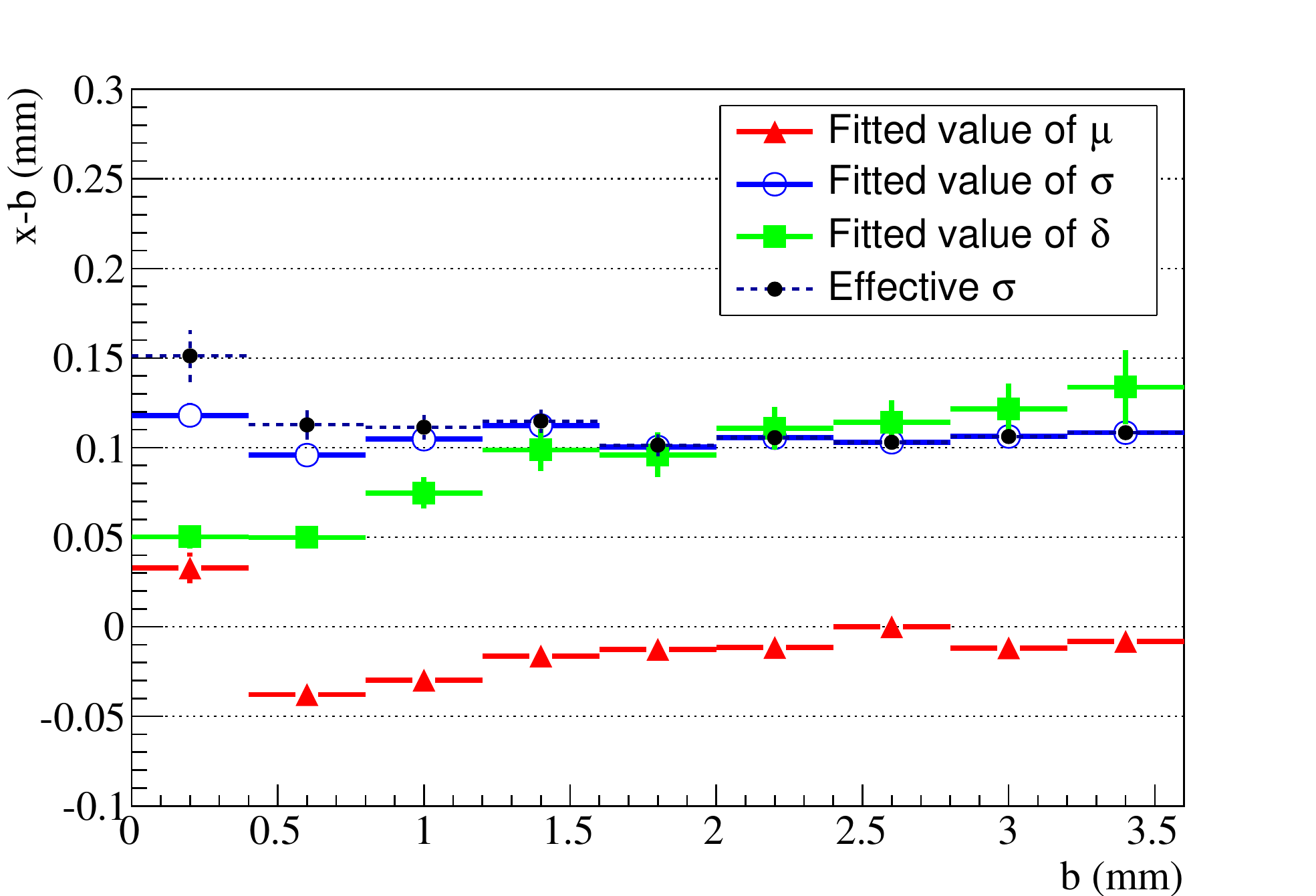}
  \label{Run7_9Slices}}
   \caption{(a) Distribution of the difference between the impact parameters returned by the drift chamber and the telescope,
  fitted by the Gaussian/exponential function discussed in the text. (b) Distribution of the fitting parameters $\mu,\,\sigma,\,\delta$ and $\sigma_\mathrm{eff}$ as functions of the impact parameter interval as defined in full text. }
 \end{figure}

Figure \ref{allreso} shows the distribution of the differences between the impact parameters returned by the drift chamber and the telescope integrated on all the impact parameters. The parameters returned by the fit are
\begin{equation*}
\begin{split}
\mu &= -0.015 \pm 0.002~\mathrm{mm}\\
\sigma &=\,\,\,\,0.106 \pm 0.002~\mathrm{mm}\\
\delta &=\,\,\,\,0.078 \pm 0.003~\mathrm{mm}
\end{split}
\end{equation*}
For a model independent comparison we compute the Full Width at Half Maximum:
\begin{equation}
\mathrm{FWHM} = \sigma \sqrt{2\mathrm{log}2} \,\,+\,\, \delta\left(\frac{\sigma^2}{\delta^2}\mathrm{log}2 + \frac{1}{2}\right) = 264~\mu m
\label{eq:FWHM}
\end{equation}
which is the same of a Gaussian with $\sigma\simeq 110$ $\mu$m.
Such result is in agreement with those obtained with indirect measurements, in which the single-hit resolution is underestimated by about 10~\%.

 Figure~\ref{Run7_9Slices} shows the parameters returned by the fit as a function of impact parameter. For impact parameters larger than 500~$\mu $m the mean of the Gaussian function is slightly negative. The sigma of the Gaussian is about 100~$\mu$m for every value of impact parameter. The parameter $\delta$ increases for large impact parameters, ranging from 50~$\mu$m to 150~$\mu$m, as a result of the decrease of the distribution asymmetry.
In addition we report for each impact parameter interval an effective resolution $\sigma_\mathrm{eff}\equiv \mathrm{FWHM}/(2\sqrt{2\log2})$, which corresponds to the standard deviation of a Gaussian distribution having a FWHM computed according to equation~(\ref{eq:FWHM}) in each bin.

\section{Conclusions}
\label{sec:conclusion}
We reported on the single-hit resolution measured by using three prototypes of the new cylindrical chamber of the MEG~II experiment. Despite the differences in the design and the operation mode of the different prototypes, the results are in agreement, yielding a resolution of about 110~$\mu$m. 
 In the final chamber further improvements are expected with the implementation of a wider bandwidth front end electronics allowing for the exploitation of the cluster timing technique~\cite{bib:clutime,bib:clutimeMC}, 
 aimed at reducing the contributions to the resolution due to the primary ionization fluctuations. 
We are confident, therefore, to achieve comparable or even better performance in the full-scale detector, which is now under construction and supposed to start data taking in 2017.% 

\section*{Acknowledgements}
We are grateful for the support and co-operation provided
by the technical and engineering staff of our institutes: L.~Beretta, S.~Bianucci, M.~Ceccanti, A.~Corvaglia, G.~Fausto, A.~Miccoli, A.~Orsini, G.~Petragnani, C.~Pinto, F.~Raffaelli, D.~Ruggieri, A.~Tazzioli, A.~Zullo.

\appendix

\section{Tracker relative alignment}
\label{app:align}
In order to compare cosmic ray tracks with drift distances, measured with respect to the anode wires, for each cell we define a Reference Frame (RF) $(x^\prime y^\prime z^\prime)$, with the $z^\prime$ axis lying along the anode wire of the cell.
\begin{figure}[!t]
  \centering
   \includegraphics[width=.62\textwidth]{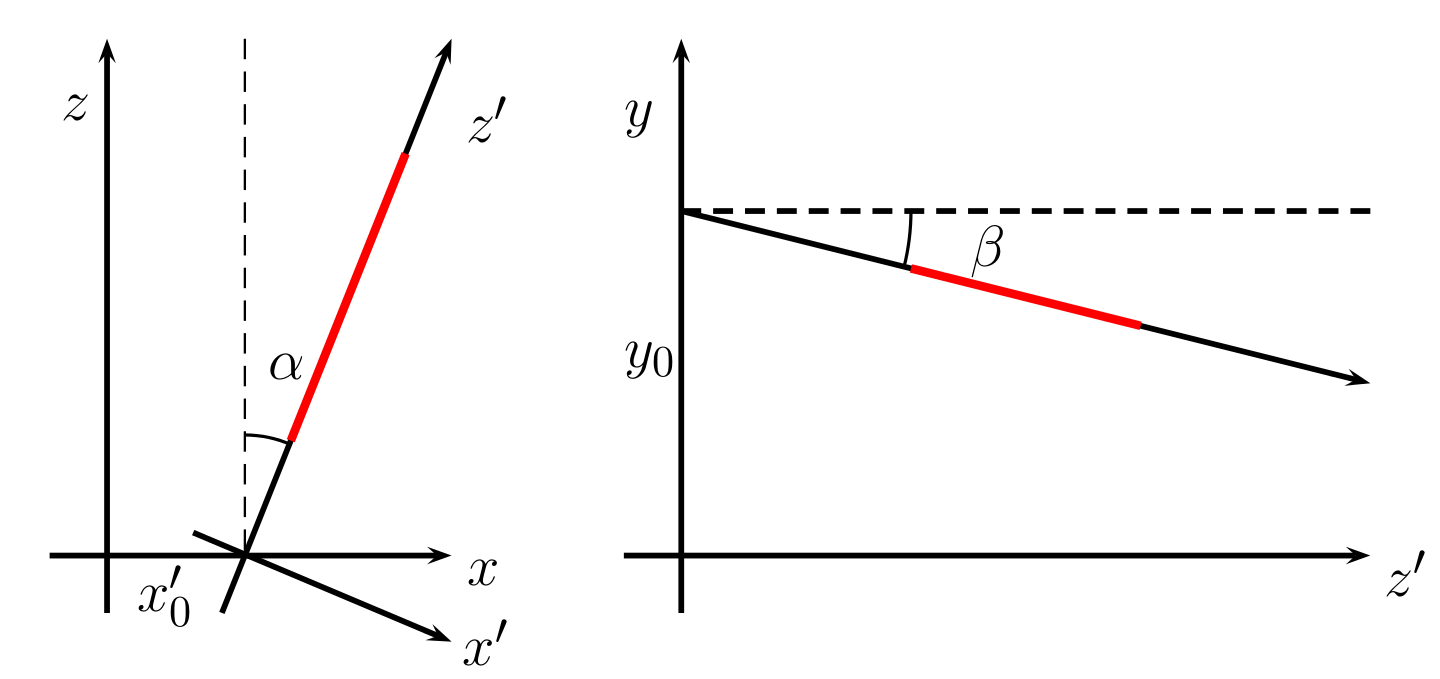} 
  \caption{Definition of alignment parameters in a horizontal (left) and vertical plane (right). The red thick lines indicate the wire position.}
  \label{align}
 \end{figure}
The coordinate transformation from one RF to the other is determined by the four parameters which identify the positions of anodes in the telescope RF (see figure~\ref{align}): an offset and angle $(x_0, \,\alpha)$ in the $xz$ plane (horizontal) and an offset and an angle $(y_0, \,\beta)$ in the $yz'$ plane (vertical).	

\begin{figure}[!b]
  \centering
\subfigure[]{\includegraphics[height=.35\textwidth]{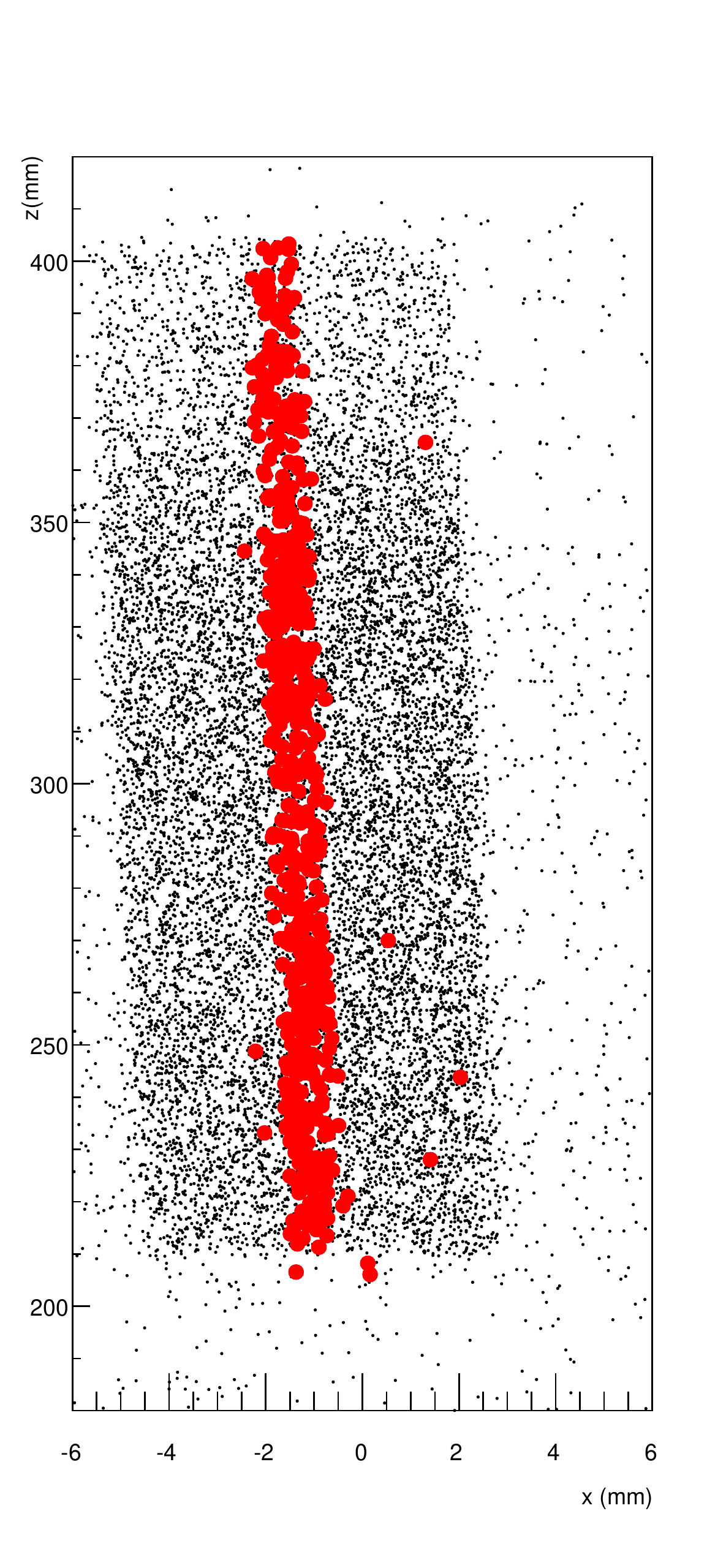}  \label{alignxa}}  \qquad 
 \subfigure[]{\includegraphics[height=.35\textwidth]{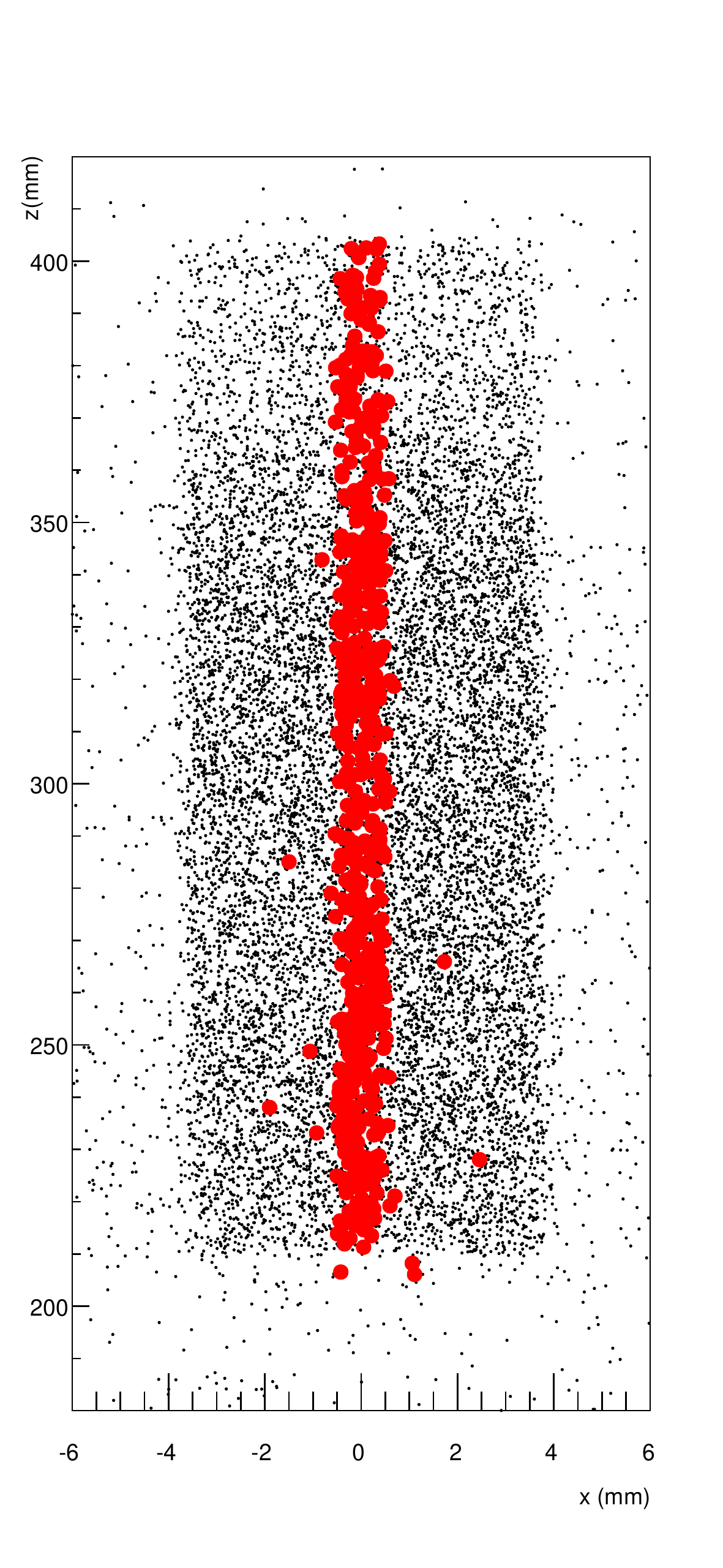} \label{alignxb}}  \qquad 
 \subfigure[]{\includegraphics[height=.32\textwidth]{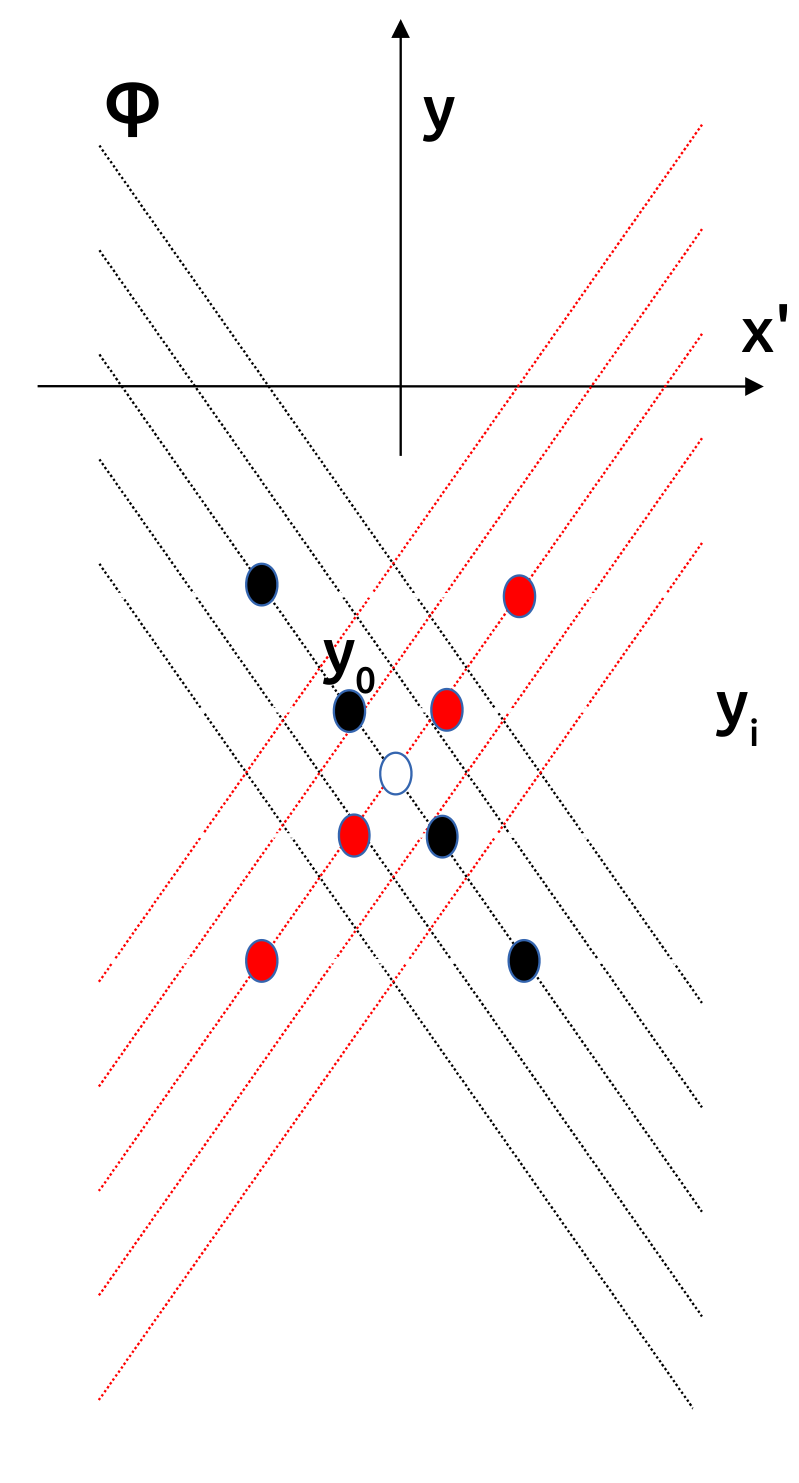}        \label{alignya}}  \qquad 
 \subfigure[]{\includegraphics[height=.32\textwidth]{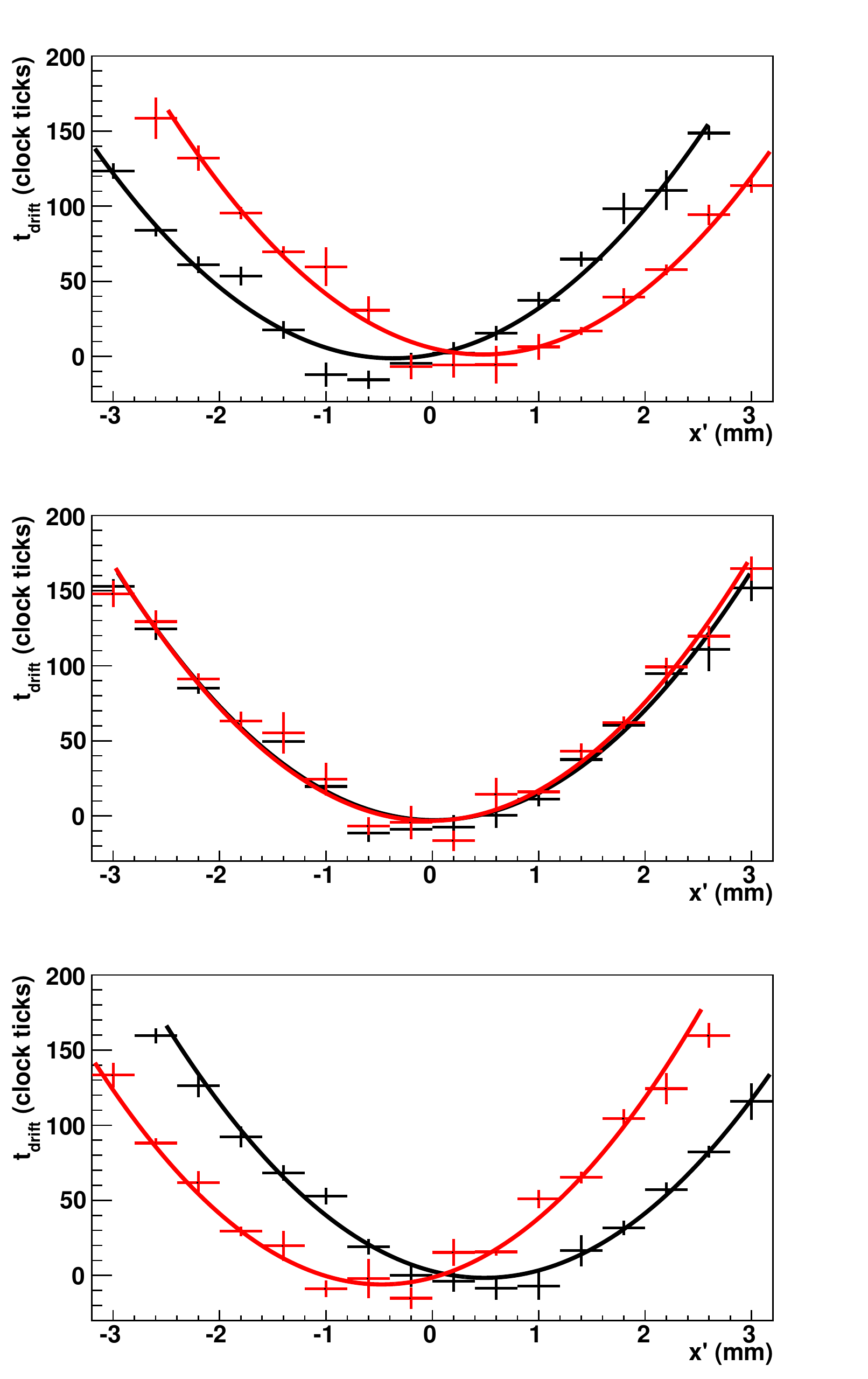} \label{alignyb}}
  \caption{Intersections of tracks at the nominal $y_\mathrm{nom}$ position of wires before (a) and after (b) the alignment procedure. Red points indicate tracks associated to drift times smaller than 5 ns. Technique used for the alignment in the $xz$ plane. Black curves refer to tracks with positive $\varphi$, red curve to those with negative $\varphi$.  (c) Effect of vertical misalignment on slanted tracks. (d) time-to-distance relations obtained assuming three different values of $y_i$ around the nominal position $y_\mathrm{nom}$ (from top to bottom $y_\mathrm{nom}+20~$mm, $y_\mathrm{nom}$, $y_\mathrm{nom}-20$~mm).}

 \end{figure}

The determination of the wire position in the $xz$ plane is obtained by calculating the track coordinates at the nominal value of the sense wire vertical coordinate $y_\mathrm{nom}$, which is measured with a ruler with a precision of about 10 mm. In this step a misplacement on the $y$ coordinate produces in fact a second-order effect since the tracks can be considered vertical to a good approximation. The crossing points of tracks corresponding to small drift times ($t<5~$ns) cluster into a region centred along the wire position (see figure~\ref{alignxa}-\ref{alignxb}). Wire parameters are obtained by fitting these data to a straight line. The misalignment angles are of the order of ten milliradians. After few iterations the uncertainty on the position of the wire is about 150 $\mu$m in the region of interest.

The alignment in the $yz$ plane is more complex because of the geometry of the apparatus: the altitude parameter $y$ is degenerate for vertical cosmic rays and, as visible in figure~\ref{phidistrib}, we have a small spread on the impinging angle around the vertical direction. Our alignment procedure is based on the effect of a vertical misalignment on slanted tracks (see figure \ref{alignya}), which produces two different $x^\prime$ corresponding to zero drift time for tracks coming from opposite angles. For maximising the effect we select offline the tracks reconstructed by the telescope with the largest angles in our acceptance (80 mrad $<|\varphi|<100$ mrad). The value of the $y_0$ of the sense wire is found in a few steps.
We consider ten different positions $y_i$ around the nominal $y_\mathrm{nom}$ at fixed $z$.
For each value $y_i$ we determine the $x^\prime_i$ associated to zero drift time separately for tracks with positive and negative slopes. For obtaining the $x^\prime_i$ related to zero drift times, we plot the measured drift times as a function of the impact parameter returned by the telescope, and a parabolic fit is performed  for obtaining the minimum of the parabola, corresponding to zero drift time (see figure \ref{alignyb}).
 The $(x^\prime_i,y_i)$ pairs are fitted to two lines (according to the slope sign) and the crossing point is chosen as the $y$ of the wire.
The measurement of the tilt angle $\beta$ is obtained by repeating this procedure in different bins of~$z$. The $y_0$ values that are obtained are almost the same for all the $z$ bins and therefore we choose a null value for $\beta$. This procedure provides the vertical coordinate of the wire with an uncertainty of about 300~$\mu$m.


\begin{thebibliography}{99}
%
\bibitem{bib:proposal} A. M. Baldini et al., {\em MEG Upgrade Proposal},  2013, \href{
http://arxiv.org/abs/1301.7225
}{ ArXiv:1301.7225}
%
\bibitem{bib:preamp} G. Chiarello et al., {\em A full front end chain for drift chambers,
\newblock  Nucl. Phys. B (Proc. Suppl.)}, \textbf{248-250} (2014) 140--142
%
\bibitem{bib:drs4} S.~Ritt, R.~ Dinapoli and U.~ Hartmann, {\em Application of the DRS chip for fast waveform digitizing, 
\newblock Nucl. Inst. Meth. }\textbf{A 623} (2010) 486--488 
%
\bibitem{bib:drs4evboard} DRS4 Evaluation Board, \href{
https://www.psi.ch/drs/evaluation-board
}{https://www.psi.ch/drs/evaluation-board}
%
\bibitem{bib:megdet} J.~Adam et al., {\em The MEG detector for $\mu^+ \rightarrow e^+\gamma$ decay search,
 \newblock  Eur.Phys.J.} \textbf{C 73} (2013) 4, 2365 
%
\bibitem{bib:btf} DAFNE Beam-Test Facility, \href{
http://www.lnf.infn.it/acceleratori/btf/}{http://www.lnf.infn.it/acceleratori/btf/}
%
\bibitem{bib:garfield} Garfield - simulation of gaseous detectors, \href{
http://garfield.web.cern.ch}{http://garfield.web.cern.ch}
%
\bibitem{bib:garfieldpp} Garfield++ - simulation of tracking detectors, \href{
http://garfieldpp.web.cern.ch/garfieldpp/}{http://garfieldpp.web.cern.ch/garfieldpp/}
%
\bibitem{bib:oldtricell} M.~Venturini,\textit{ Ageing and performance studies of drift chamber prototypes for the MEG II experiment, Nuovo Cim. }\textbf{C 38} (2015) no.1, 22
%
\bibitem{bib:geant4} S.~Agostinelli et al., {\em Geant4 - a simulation toolkit, 
\newblock Nucl. Inst. Meth}. \textbf{A 506} (2003) 250--303
%
\bibitem{bib:telescopio} L. Galli et al., {\em A silicon based cosmic ray telescope as an external
tracker to measure detector performance,
\newblock  IEEE Trans. Nucl. Sci.} \textbf{62} (2015) 1, 395--402
%
\bibitem{bib:tricell} M.~Venturini et al., {\em MEG II drift chamber characterization with the silicon based cosmic ray tracker at INFN Pisa,
\newblock Nucl. Inst. Meth.} \textbf{A 824} (2016) 595--597
%
\bibitem{bib:babar} B. Aubert et al. (Babar Collaboration), {\em The BaBar Detector, Nucl. Inst. Meth.} \textbf{A 479} (2002) 1--116
%
\bibitem{bib:clutime} F. Grancagnolo et al., {\em Cluster counting in helium based gas mixtures, 
\newblock Nucl. Inst. Meth.} \textbf{A 386} (1997) 458--469
%
\bibitem{bib:clutimeMC}
G.~Signorelli, A.~D'Onofrio, and M.~Venturini,
\newblock {\em A novel method to estimate the impact parameter on a drift cell by
  using the information of single ionization clusters}, 
\newblock { \em Nucl. Instrum. Meth.}, \textbf{A 824} (2016) 581--583

\end{thebibliography}
\end{document}